\documentstyle[12pt]{article}
\newcommand{\nc}{\newcommand}
\nc{\renc}{\renewcommand}

\renc{\baselinestretch}{1.15}
\nc{\com}[1]{\ \\ \ {\bf \# {#1}}\\ \ }
\nc{\bort}[1]{}
\newlength{\overeqskip}
\newlength{\undereqskip}
\nc{\be}[1]{\begin{equation} \mbox{$\label{#1}$}}
\nc{\bea}[1]{\begin{eqnarray} \mbox{$\label{#1}$}}
\nc{\Section}[2]{\section{\sc #2}\label{#1}}
\nc{\Subsection}[2]{\subsection{\sc #2}\label{#1}}
\nc{\Bibitem}[1]{\bibitem{#1}}
\nc{\Label}[1]{\label{#1}}

\nc{\eea}{\vspace{\undereqskip}\end{eqnarray}}
\nc{\ee}{\vspace{\undereqskip}\end{equation}}
\nc{\bdm}{\begin{displaymath}}
\nc{\edm}{\end{displaymath}}
\nc{\dpsty}{\displaystyle}
\nc{\bc}{\begin{center}}
\nc{\ec}{\end{center}}
\nc{\ba}{\begin{array}}
\nc{\ea}{\end{array}}
\nc{\bab}{\begin{abstract}}
\nc{\eab}{\end{abstract}}
\nc{\btab}{\begin{tabular}}
\nc{\etab}{\end{tabular}}
\nc{\bit}{\begin{itemize}}
\nc{\eit}{\end{itemize}}
\nc{\ben}{\begin{enumerate}}
\nc{\een}{\end{enumerate}}
\nc{\bfig}{\begin{figure}}
\nc{\efig}{\end{figure}}
\nc{\seqnoll}{\setcounter{equation}{0}}
\renc{\theequation}{\thesection.\arabic{equation}}
\renc{\Section}[2]{\bc\section{\sc #2}\Label{#1}\seqnoll\ec}

\nc{\refc}[1]{\mbox{Ref.~\cite{#1}}}
\nc{\refs}[1]{\mbox{Refs.~\cite{#1}}}

\nc{\eqs}[2]{\mbox{Eqs.~(\ref{#1},\,\ref{#2})}}
\nc{\eq}[1]{\mbox{Eq.~(\ref{#1})}}
\nc{\rf}[1]{(\ref{#1})}

\nc{\figs}[2]{\mbox{Figs.~\ref{#1} and \ref{#2}}}
\nc{\fig}[1]{\mbox{Fig.~\ref{#1}}}

\nc{\figcap}[1]{\refstepcounter{figure}
        {\bf Figure \thefigure}: {\small\sl #1}}
\nc{\tabcap}[1]{\refstepcounter{table}
        {\bf Table \thetable}: {\small\sl #1}}

\def\jump{\vskip 1truecm\noindent}

\nc{\tag}[1]{\label{#1} \marginpar{{\footnotesize #1}}}
\nc{\mtag}[1]{\label{#1} \mbox{\marginpar{{\footnotesize #1}}}}

\nc{\etal}{\mbox{\it et al. }}
\nc{\ie}{{\it i.e.}}
\nc{\eg}{{\it e.g.}}

\nc{\arreq}{&\!\!\!=\!\!\!&}
\nc{\arrmi}{&\!\!\!!-\!\!\!&}
\nc{\arrpl}{&\!\!\!+\!\!\!&}
\nc{\arrap}{&\!\!\!\approx\!\!\!&}
\nc{\non}{\nonumber}
\nc{\align}{\!\!\!\!\!\!\!\!&&}
\nc{\del}{\partial}
\nc{\mat}[4]{{\left(\ba{cc} #1 & #2 \\ #3 & #4 \ea\right)}}

\def\simleq{\; \raise0.3ex\hbox{$<$\kern-0.75em
      \raise-1.1ex\hbox{$\sim$}}\; }
\def\simgeq{\; \raise0.3ex\hbox{$>$\kern-0.75em
      \raise-1.1ex\hbox{$\sim$}}\; }
\nc{\DOT}{\hspace{-0.08in}{\bf .}\hspace{0.1in}}
\nc{\Laada}{\hbox {$\sqcap$ \kern -1em $\sqcup$}}
\nc\loota{{\scriptstyle\sqcap\kern-0.55em\hbox{$\scriptstyle\sqcup$}}}
\nc\Loota{{\sqcap\kern-0.65em\hbox{$\sqcup$}}}
\nc\laada{\Loota}
\nc{\qed}{\hskip 3em \hbox{\BOX} \vskip 2ex}
\def\Re{{\rm Re}\hskip2pt}
\def\Im{{\rm Im}\hskip2pt}
\nc{\real}{{\rm I \! R}}
\nc{\Z}{{\sf Z \!\!\! Z}}
\nc{\complex}{{\rm C\!\!\! {\sf I}\,\,}}
\def\bigid{\leavevmode\hbox{\small1\kern-3.8pt\normalsize1}}
\def\id{\leavevmode\hbox{\small1\kern-3.3pt\normalsize1}}
\nc{\slask}{\!\!\!\!/}
\nc{\sslask}{\!\!\!/}
\nc{\bis}{{\prime\prime}}
\nc{\pa}{\partial}
\nc{\na}{\nabla}
\nc{\ra}{\rangle}
\nc{\la}{\langle}
\nc{\goto}{\rightarrow}
\nc{\swap}{\leftrightarrow}

\nc{\EE}[1]{ \mbox{$\cdot10^{#1}$} }
\nc{\abs}[1]{\left|#1\right|}
\nc{\at}[2]{\left.#1\right|_{#2}}
\nc{\norm}[1]{\|#1\|}
\nc{\abscut}[2]{\abs{#1}_{\scriptscriptstyle#2}}
\nc{\vek}[1]{{\rm\bf #1}}
\nc{\integral}[2]{\int\limits_{#1}^{#2}}
\nc{\inv}[1]{\frac{1}{#1}}
\nc{\dd}[2]{{{\partial #1}\over{\partial #2}}}
\nc{\ddd}[2]{{{{\partial}^2 #1}\over{\partial {#2}^2}}}
\nc{\dddd}[3]{{{{\partial}^2 #1}\over
        {\partial #2 \partial #3}}}
\nc{\dder}[2]{{{d #1}\over{d #2}}}
\nc{\ddder}[2]{{{d^2 #1}\over{d {#2}^2}}}
\nc{\dddder}[3]{{d^2 #1}\over
        {d #2 d #3}}
\nc{\dbar}{d \hspace{-0.25em} \raisebox{0.45ex}{\Large -}}
\nc{\cc}{\mbox{$c.c.$ }}
\nc{\hc}{\mbox{$h.c.$ }}
\nc{\cf}{cf.\ }
\nc{\erfc}{{\rm erfc}}
\nc{\Tr}{{\rm Tr\,}}
\nc{\tr}{{\rm tr\,}}
\nc{\pol}{{\rm pol}}
\nc{\sign}{{\rm sign}}
\nc{\bfT}{{\bf T }}

\nc{\cA}{{\cal A}}
\nc{\cB}{{\cal B}}
\nc{\cD}{{\cal D}}
\nc{\cE}{{\cal E}}
\nc{\cF}{{\cal F}}
\nc{\cG}{{\cal G}}
\nc{\cH}{{\cal H}}
\nc{\cL}{{\cal L}}
\nc{\cM}{{\cal M}}
\nc{\cO}{{\cal O}}
\nc{\cT}{{\cal T}}
\nc{\rvac}[1]{|{\cal O}#1\rangle}
\nc{\levac}[1]{\langle{\cal O}#1|}
\nc{\rvacb}[1]{|{\cal O}_\beta #1\rangle}
\nc{\levacb}[1]{\langle{\cal O}_\beta #1 |}
\nc{\bb}{\bar{\beta}}
\nc{\ctH}{\tilde{\cal H}}
\nc{\chH}{\hat{\cal H}}
\nc{\erf}{ {\rm erf}}
\nc{\aae}{\"{a}}
\nc{\eo}{\"{o}}
\nc{\Ea}{\"{A}}
\nc{\Eo}{\"{O}}
\nc{\al}{\alpha}
\nc{\Del}{\Delta}
\nc{\e}{\epsilon}
\nc{\eps}{\varepsilon}
\nc{\lam}{\lambda}
\nc{\om}{\omega}
\nc{\Om}{\Omega}
\nc{\mn}{{\mu\nu}}
\nc{\k}{\kappa}
\nc{\vp}{\varphi}
\nc{\pub}[4]{\Bibitem{#1}#2, {\sl ``#3''}, #4.}
\nc{\acta}[3]{{\it Acta Phys.~Austr.\ }{{\bf #1}}{(#2)}{#3}}
\nc{\advp}[3]{{\it  Adv.\ in\ Phys.\ }{{\bf #1} {(#2)} {#3}}}
\nc{\annp}[3]{{\it  Ann.\ Phys.\ (N.Y.)\ }{{\bf #1} {(#2)} {#3}}}
\nc{\apl}[3]{{\it  Appl. Phys. Lett. }{{\bf #1} {(#2)} {#3}}}
\nc{\apj}[3]{{\it  Ap.\ J.\ }{{\bf #1} {(#2)} {#3}}}
\nc{\apjl}[3]{{\it  Ap.\ J.\ Lett.\ }{{\bf #1} {(#2)} {#3}}}
\nc{\app}[3]{{\it Astropart.\ Phys.\ }{{\bf #1} {(#2)} {#3}}}
\nc{\cmp}[3]{{\it  Comm.\ Math.\ Phys.\ }{{ \bf #1} {(#2)} {#3}}}
\nc{\cqg}[3]{{\it  Class.\ Quant.\ Grav.\ }{{\bf #1} {(#2)} {#3}}}
\nc{\epl}[3]{{\it  Europhys.\ Lett.\ }{{\bf #1} {(#2)} {#3}}}
\nc{\ijmp}[3]{{\it Int.\ J.\ Mod.\ Phys.\ }{{\bf #1} {(#2)} {#3}}}
\nc{\ijtp}[3]{{\it Int.\ J.\ Theor.\ Phys.\ }{{\bf #1} {(#2)} {#3}}}
\nc{\jmp}[3]{{\it  J.\ Math.\ Phys.\ }{{ \bf #1} {(#2)} {#3}}}
\nc{\jpa}[3]{{\it  J.\ Phys.\ A\ }{{\bf #1} {(#2)} {#3}}}
\nc{\jpc}[3]{{\it  J.\ Phys.\ C\ }{{\bf #1} {(#2)} {#3}}}
\nc{\jpg}[3]{{\it J.~Phys.~G:~Nucl.~Part.~Phys.~}{{\bf #1} {(#2)} {#3}}}
\nc{\jap}[3]{{\it J.\ Appl.\ Phys.\ }{{\bf #1} {(#2)} {#3}}}
\nc{\jpsj}[3]{{\it J.\ Phys.\ Soc.\ Japan\ }{{\bf #1} {(#2)} {#3}}}
\nc{\lmp}[3]{{\it Lett.\ Math.\ Phys.\ }{{\bf #1} {(#2)} {#3}}}
\nc{\lncim}[3]{{\it  Lett.\ Nuov.\ Cim.\ }{{\bf #1} {(#2)} {#3}}}
\nc{\mpl}[3]{{\it  Mod.\ Phys.\ Lett.\ }{{\bf #1} {(#2)} {#3}}}
\nc{\ncim}[3]{{\it  Nuov.\ Cim.\ }{{\bf #1} {(#2)} {#3}}}
\nc{\np}[3]{{\it  Nucl.\ Phys.\ }{{\bf #1} {(#2)} {#3}}}
\nc{\pr}[3]{{\it Phys.\ Rev.\ }{{\bf #1} {(#2)} {#3}}}
\nc{\pra}[3]{{\it  Phys.\ Rev.\ }{{\bf A#1} {(#2)} {#3}}}
\nc{\prb}[3]{{\it  Phys.\ Rev.\ }{{\bf B#1} {(#2)} {#3}}}
\nc{\prc}[3]{{\it  Phys.\ Rev.\ }{{\bf C#1} {(#2)} {#3}}}
\nc{\prd}[3]{{\it  Phys.\ Rev.\ }{{\bf D#1} {(#2)} {#3}}}
\nc{\prl}[3]{{\it Phys.\ Rev.\ Lett.\ }{{\bf #1} {(#2)} {#3}}}
\nc{\pl}[3]{{\it  Phys.\ Lett.\ }{{\bf #1} {(#2)} {#3}}}
\nc{\prep}[3]{{\it Phys\. Rep.\ }{{\bf #1} {(#2)} {#3}}}
\nc{\prsl}[3]{{\it Proc.\ R.\ Soc.\ London\ }{{\bf #1} {(#2)} {#3}}}
\nc{\ptp}[3]{{\it  Prog.\ Theor.\ Phys.\ }{{\bf #1} {(#2)} {#3}}}
\nc{\ptps}[3]{{\it  Prog\ Theor.\ Phys.\ suppl.\ }{{\bf #1} {(#2)} {#3}}}
\nc{\ppnp}[3]{{\it  Prog.\ Nucl.\ Part.\ Phys.\ }{{\bf #1} {(#2)} {#3}}}
\nc{\physa}[3]{{\it  Physica\ A\ }{{\bf #1} {(#2)} {#3}}}
\nc{\physb}[3]{{\it  Physica\ B\ }{{\bf #1} {(#2)} {#3}}}
\nc{\phys}[3]{{\it Physica\ }{{\bf #1} {(#2)} {#3}}}
\nc{\rmp}[3]{{\it  Rev.\ Mod.\ Phys.\ }{{\bf #1} {(#2)} {#3}}}
\nc{\rpp}[3]{{\it Rep.\ Prog.\ Phys.\ }{{\bf #1} {(#2)} {#3}}}
\nc{\sjnp}[3]{{\it Sov.\ J.\ Nucl.\ Phys.\ }{{\bf #1} {(#2)} {#3}}}
\nc{\spjetp}[3]{{\it Sov.\ Phys.\ JETP\ }{{\bf #1} {(#2)} {#3}}}
\nc{\yf}[3]{{\it Yad.\ Fiz.\ }{{\bf #1} {(#2)} {#3}}}
\nc{\zetp}[3]{{\it Zh.\ Eksp.\ Teor.\ Fiz.\ }{{\bf #1} {(#2)} {#3}}}
\nc{\zp}[3]{{\it Z.\ Phys.\ }{{\bf #1} {(#2)} {#3}}}
\nc{\zpc}[3]{{\it Z.\ Phys.\ C\ }{{\bf #1} {(#2)} {#3}}}
\nc{\ibid}[3]{{\sl ibid.\ }{{\bf #1} {#2} {#3}}}
\newcommand{\minus}{\!-\!}
\newcommand{\plus}{\!+\!}
\nc{\nn}{\nonumber \\*}
\nc{\Psibar}{\overline{\Psi}}
\nc{\LL}{Landau level}
\nc{\LLL}{lowest Landau level}
\nc{\amm}{anomalous magnetic moment}
\nc{\leff}{\cL_{\rm eff}}
\nc{\leffq}{\cL_{\rm eff}^{\rm  ferm}}
\nc{\leffs}{\cL_{\rm eff}^{\rm scal}}
\nc{\leffv}{\cL_{\rm eff}^{\rm vec}}
\nc{\leffc}{\cL_{\rm eff}^{\rm QCD}}
\nc{\ltre}{\cL_{\rm tree}}
\nc{\lvac}{\cL_{\rm vac}}
\nc{\lvacq}{\cL_{\rm vac}^{\rm ferm}}
\nc{\lvacs}{\cL_{\rm vac}^{\rm scal}}
\nc{\lvacv}{\cL_{\rm vac}^{\rm vec}}
\nc{\lvacc}{\cL_{\rm vac}^{\rm QCD}}
\nc{\dlvac}{\Delta \cL_{\rm vac}^{\rm vec}}
\nc{\lmat}{\cL_{\rm mat}}
\nc{\lmatn}{\cL_{\rm mat,0}}
\nc{\lmate}{\cL_{\rm mat,1}}
\nc{\lmatq}{\cL_{\rm mat}^{\rm ferm}}
\nc{\lmatc}{\cL_{\rm mat}^{\rm QCD}}
\nc{\lmats}{\cL_{\rm mat}^{\rm scal}}
\nc{\lmatv}{\cL_{\rm mat}^{\rm vec}}
\nc{\dlmatq}{\Delta \cL_{\rm mat}^{\rm ferm}}
\nc{\dlmat}{\Delta \cL_{\rm mat}^{\rm vec}}
\nc{\mvac}{M_{\rm vac}}
\nc{\cvac}{\chi_{\rm vac}}
\nc{\mmat}{M_{\rm mat}}
\newcommand{\eeff}{e_{\rm eff}}
\nc{\geff}{g_{\rm eff}}

\newcommand{\ch}{\hat{\chi}}
\newcommand{\chc}{\hat{\chi}^{\rm QCD}}
\nc{\gef}{g_{\rm eff}}
\nc{\psibar}{\overline{\psi}}
\nc{\tp}{\tilde{\Pi}}
\nc{\pimat}{\Pi_{\rm mat}^{00}}
\nc{\pitr}{\Pi_{\rm mat}^T}
\begin{document}
%
\thispagestyle{empty}
\begin{flushright}
            G\"{o}teborg ITP 96-01, hep-ph/9601259\\
	Revised version, Mars 1996\\
 \end{flushright}
\begin{center}
{\Large\bf   ASYMPTOTIC FREEDOM FROM THERMAL\\[2ex]
	 AND VACUUM MAGNETIZATION}\\[5mm]
\normalsize
{\large David Persson\footnote{\noindent Email address:
tfedp@fy.chalmers.se}}\\
Institute of Theoretical Physics\\
   Chalmers University of Technology and G\"oteborg  University\\
 S-412 96 G\"oteborg, Sweden \\
\ec
\bort{\begin{center}
\leavevmode
\epsfbox{logo.ps}\vspace{1ex}
\end{center}}
%
%
\vfill
\bc
{\bf Abstract} \\
\ec
{\small
\begin{quotation}
\noindent
We calculate the effective Lagrangian for a  magnetic field in spinor, scalar
and vector QED. Connections are then made to $SU(N_C)$ Yang--Mills theory and
QCD. The magnetization and the corresponding effective charge are obtained 
from the effective Lagrangian. The renormalized vacuum magnetization will
depend on the renormalization scale chosen. Regardless of this, the effective
charge  decreasing  with the magnetic field, as in QCD, corresponds to  anti-
screening and asymptotic freedom. In spinor and scalar QED on the other hand,
the effective charge is increasing with the magnetic field, corresponding to 
screening. Including effects due to finite temperature and density, we 
comment on the effective charge in a degenerate fermion gas, increasing 
linearly with the chemical potential. Neglecting the tachyonic mode, we find
that in hot QCD the effective charge is {\em decreasing} as the inverse 
temperature, in favor for the formation of a quark-gluon plasma. However,
including the real part of the contribution from the tachyonic mode, we find 
instead an effective charge {\em increasing}  with the temperature. Including
a thermal gluon  mass, the effective charge  in hot QCD is group invariant
(unlike in the two cases above), and decreases logarithmically in accordance
to the vacuum renormalization group equation, with the temperature as the 
momentum scale.
\end{quotation}}
\vfill
\newpage
\normalsize
\setcounter{page}{1}
%
\Section{intro}{Introduction}
We  calculate the one loop effective Lagrangian density for a static uniform
magnetic field in different gauge theories. From this the effective charge,
the magnetization, and other relevant objects are obtained.
In a recent publication~\cite{elmforslps95},
the vacuum of spinor and scalar QED
was shown to exhibit  paramagnetic properties.
 On the other hand, Nielsen related in a well known
article \cite{nielsen81}, asymptotic freedom of QCD to a paramagnetic
vacuum of
mass-less QCD. With the same reasoning, the vacuum of Abelian gauge
theories  (like QED) should exhibit a diamagnetic behavior. In
 the present letter we resolve the ostensible discrepancy between the two
different approaches above.  This discrepancy originates in the use of
renormalized quantities in \refc{elmforslps95}, verses
 the use of bare quantities, in order to get analogy with a classical
dielectricum, in \refc{nielsen81}.  In  Section~\ref{sec-dielek} we
shall briefly review the properties of a dielectricum, and relate the
effective charge to the magnetic behavior.  In  Section~\ref{sec-efflag} we
consider the generic effective Lagrangian for a background magnetic field,
and how the magnetization is obtained from it. We also give another definition
of an
effective charge, applicable also at finite temperature and density.
 In  Section~\ref{sec-vacuum} we calculate the vacuum effective Lagrangian for
 spinor, scalar and vector QED in the massive as well as the mass-less case.
Similar calculations have earlier been performed in for example
\refc{nknielseno78} (in terms of the vacuum energy), and in \refc{kaminski81}
(in QED).\bigskip

Recently much attention has been paid to the $\beta$-function and the effective
charge in QCD at high temperature ( see e.g.
\refs{nakkagawany88,landsman89,eijcksw94,elmforsk95,chaichianh95}), and
high density (see e.g. \refc{kapusta79}), motivated by the suggested
formation of a quark--gluon plasma under such circumstances~\cite{collinsp75}.
 The results differ depending on
the  different choices
of gauge, gauge fixing parameter~\cite{eijcksw94,elmforsk95},
  and on the vertex considered~\cite{nakkagawany88}.
The effective charge obtained from the effective  Lagrangian of a background
field should be free of these diseases. In the background field formalism,
as used also in \refs{landsman89,eijcksw94,elmforsk95,chaichianh95}, the
renormalized couplings of the different vertices are kept equal and related to
the renormalization of the background field. We
 shall in Section~\ref{sec-thermal} calculate the effective Lagrangian in the
presence of a heat and charge bath, and relate it to the free energy of the
 plasma. Summing only over the physical degrees of freedom, this should be
gauge independent. However, to perform the explicit calculations we have
chosen a certain gauge for the background magnetic field, and for the
mass-less gluon field. Effective
Lagrangians for a static uniform magnetic field in a thermal environment
have been considered earlier. For scalar and spinor QED in for example
\refs{dittrich79,elmforsps93}, and in $SU(N_c)$ Yang--Mills theories
(for $N_c=2,3$) in for example
\refs{kapusta81,mullerr81,dittrichs81,ninomiyas81,leonidov91,starinetsvz94}.
However,  in the Yang--Mills
theories, they   focused
on the effective Lagrangian (or equivalently the thermodynamic pressure)
in order to  investigate a  possible phase transition, and
we disagree with some of the previous results.
Since perturbation theory in this situation  actually only is valid
after the phase transition to a quark--gluon plasma has taken place,
we shall here
mainly focus on the effective charge obtained from the effective Lagrangian.
A corresponding  effective charge has earlier been considered in
QED~\cite{chodosos88,elmforsps93}, but
to our knowledge not in non-Abelian gauge theories.

We shall use a naive real-time formalism,
valid here since there are no propagators with coinciding momenta, and
calculate the thermal effective Lagrangian in spinor, scalar and vector QED.
$SU(N_c)$ Yang-Mills theory is then related to charged
 mass-less vector bosons, and
connections are  made with QCD in Section~\ref{sec-qcd}. When possible we
compare
with previous results, and make some corrections. We also consider the effects
when
a thermal gluon mass is taken into account.
Finally we discuss the results here obtained, and their relevance
 in Section~\ref{sec-disc}.

 Since the topics here considered provide an extremely nice and simple
example of renormalization, we shall be fairly explicit.

\Section{sec-dielek}{Magnetization and (Anti-) Screening}
In his very pedagogical example, Nielsen\cite{nielsen81} described
asymptotic freedom (anti-screening) in QCD, in the same way as the intuitively
clear picture of screening in an ordinary dielectric medium. Let us first
recapitulate some of the basic features of a classical homogeneous
dielectricum. The effective coupling of two test charges separated by a
distance $r$ is
\be{epseff}
 e^2_\varepsilon(r)=\frac{e^2_0}{\varepsilon(r)}~~~,
\ee
where  $e_0$ is the undressed charge measured in
the absence of the medium. The dielectric permittivity $\varepsilon(r)$
must approach unity as $r \rightarrow 0$, since then there is no shielding
medium between the two test charges.  Screening means that the medium will
be polarized around the test charge, so that the effective charge measured at
$r>0$, will be smaller than the undressed charge (at $r=0$). In terms of
the dielectric permittivity we thus have $\varepsilon>1$. If, on the other
hand, the polarization of the medium is such that the effective charge is
larger than the undressed charge we have anti-screening, and correspondingly
$\varepsilon <1$.
\bigskip

In analogy to this, Nielsen  calculated the effective charge in QCD, with
the dielectricum consisting of the quantum mechanical vacuum, containing
virtual particles only. In a quantum field theory one will encounter
divergences that after a proper regularization may be removed by a rescaling
of the parameters appearing in the  original Lagrangian (renormalization).
In a classical theory there are  no such divergences, so in order to get full
analogy with the example of a dielectric medium, Nielsen used cut-off
regularized bare (i.e. before renormalization) quantities. An ultraviolet
momentum cut-off $\Lambda$, corresponds to a smallest distance $r_0=1/\Lambda$,
where the bare charge is measured, corresponding to the classical
undressed charge, at $r=0$.

The relativistic invariance of the vacuum of a quantum field theory requires
that the permittivity is connected to the magnetic permeability
$\mu_{\rm perm}$,
through
\be{epsmu}
	\varepsilon \mu_{\rm perm} =1~~~.
\ee
This has no counterpart in an ordinary polarizable medium. It turns out that
 it is easier to calculate the magnetic susceptibility $\chi$ (such that
$\mu_{\rm perm} =1+\chi$) in a background (color) magnetic field $B$, than to
directly calculate the permittivity in a (color) electric field.
By a heuristic reasoning, Nielsen related the distance to the field
strength, according to $r \approx 1/\sqrt{eB}$ (we are assuming $eB>0$).
We thus find the dielectric permittivity
\be{epsmag}
\varepsilon(r)= \left.\frac1{1+\chi(B)} \right|_{eB\rightarrow 1/r^2}~~~.
\ee
In terms of bare regularized quantities (denoted by the subscript ``$b$''), we
thus find that screening  ($\varepsilon_b>1$),  corresponds to $\chi_b<0$,
i.e. diamagnetism; and anti-screening ($\varepsilon_b<1$),  corresponds to
$\chi_b>0$,
i.e.  paramagnetism.

How could then possibly the Abelian gauge theories discussed in
\refc{elmforslps95}
exhibit a paramagnetic behavior, when they cannot be asymptotically free?

The answer to this lies in the renormalization procedure. We can never measure
such things as bare quantities, but renormalization is necessary in order to
obtain the physical parameters. The renormalization is performed at some
momentum scale $\lambda$, arbitrary but finite.
 The reference charge will
 be the renormalized charge, measured at momentum scale $\lambda$, or
 equivalently on a distance $r_1 \approx 1/\lambda$. For the clarity of the
reasoning, and the comparison with the classical dielectricum, we are
assuming some sort of on--shell renormalization, where the renormalized
charge corresponds to the charge measured at the renormalization scale.
In general it is sufficient only to remove the divergences in the
renormalization procedure. Then one will have some finite relation between the
physical charge actually measured, and the renormalized charge that just
is a parameter of the theory.
In analogy to the classical dielectricum we thus have
\be{alphaeps}
  	e^2_\varepsilon(r)= \frac{e^2_1}{\varepsilon_1(r)}~~~,
\ee
where $e^2_1$ is the charge measured at distance $r_1$, i.e.
$\varepsilon_1(r_1)=1$, corresponding to the charge  renormalized at momentum
$\lambda_1 \approx 1/r_1$.
The general criteria for screening is that the effective charge
should decrease with the distance or increase with the momentum scale, and
vice versa for anti-screening. When considering distances smaller than $r_1$,
screening  instead corresponds to $\varepsilon_1<1$, and anti-screening
to $\varepsilon_1>1$. Without knowledge of the scale at which the reference
charge is measured, the magnitude of $\varepsilon$, will thus not tell
whether we have screening or anti-screening.

In the renormalized spinor and scalar QED
 considered in \refc{elmforslps95} the
renormalization
is performed at the lowest momentum scale (as is standard),
 corresponding to $r_1\rightarrow
\infty$. In this case we thus always have $r<r_1$, and  screening here
corresponds to $\varepsilon<1$, which means paramagnetism $\chi>0$, and
similarly anti-screening would  correspond to $\varepsilon>1$,
and thereby  $\chi<0$, i.e.  diamagnetism. However, if the theory exhibits
anti-screening (like QCD), the effective charge is becoming infinitely large
at large distances, so the renormalization cannot be performed at vanishing
momentum scale.

\Section{sec-efflag}{Effective  Lagrangian and Effective Coupling}
The generic effective Lagrangian for a background magnetic field may be
separated
 into its different contributions
\be{efflagsep}
\leff= \cL_{\rm tree} + \lvac +\lmat~~~.
\ee
Here
\be{ltree}
	\cL_{tree}= - B^2/2\qquad,
\ee
is the free tree level part; $ \lvac$ is the vacuum contribution due to virtual
particles; and $\lmat$ is the contribution due to real particles at finite
temperature
and density. The vacuum contribution $ \lvac$ is calculated to one-loop order
in
Section~\ref{sec-vacuum}, and the
matter contribution $\lmat$ will be considered in
Section~\ref{sec-thermal}.

{}From the effective Lagrangian we may obtain the magnetization
\be{magdef}
	M \equiv \mvac+ \mmat = \frac{\partial }{\partial B}(\lvac +\lmat)~~~.
\ee
The vacuum magnetization, originally proposed in  \refc{elmforslps95}, is a
real
physical quantity, but only gives measurable effects at extremely high field
strengths.
Performing another derivative we find the magnetic susceptibility
\be{suscdef}
	\chi \equiv \cvac + \chi_{\rm mat} =
	 \frac{\partial }{\partial B}(\mvac +\mmat)~~~.
\ee
At vanishing temperature and density we may use the vacuum magnetic
susceptibility to
obtain the effective charge according to \eq{epsmag} and \eq{alphaeps}.
In the presence of matter at finite temperature ($T$) and chemical
potential ($\mu$), Lorentz invariance is broken  when choosing a preferred
frame
 of reference in which the medium is at rest and in equilibrium.
Then there is no connection between the permittivity
and the permeability. We must therefore find another way of obtaining the
effective charge in this case.
It may be done  through the identification~\cite{chodosos88}
\be{effeid}
 	\leff \approx -\frac12 B_{\rm eff}^2 =
	-\frac12 \frac{(eB)^2}{e_{\rm eff}^2}~~~,
\ee
where we have  used that  $eB=e_{\rm eff} B_{\rm eff}$ is required to be
invariant
under renormalization.
Performing the derivative with
respect to $(eB)^2$, that is invariant under renormalization as well as
Lorentz transformations ($B^2-E^2$ is Lorentz invariant and here reduced to
$B^2$), we find
\be{aeffdef}
	\frac1{e_{\rm eff}} \equiv- 2 \frac{\del \leff}{\del
	(eB)^2}
	=\frac1{e^2}-\frac1{eB}\frac{M}{e}~~~,
\ee
where $M$ is the magnetization. Notice that one may also define the magnetic
susceptibility as the response function of the magnetization $M=\chi B$. In
this case
we find
\be{altadef}
	\frac1{e_{\varepsilon}^2}=\frac1{e^2} \frac1{1+\chi} \simeq
	\frac1{e^2}-\frac1{eB}\frac{M}{e}~~~,
\ee
to the lowest order in the coupling.\bigskip

 An external magnetic field $H$ is
 introduced by adding a term $\cL_{\rm ext}=
j^\nu_{\rm ext}A_{\nu}$ to the effective Lagrangian. Neglecting a surface
term we find $ \cL_{\rm ext}= {\bf B} \cdot {\bf H}$.
By construction $\leff$ is  invariant
under renormalization. If $\leff + \cL_{\rm ext}$ is to be invariant, the
invariance of $eB$ requires $H/e$ to be invariant. This seems quite reasonable,
 since $\nabla \times  {\bf H}= {\bf j}_{\rm ext}$. The external field $H$ is
thus only a function of the external charges, and thus should scale as a
charge. Minimizing $\leff$ with respect to $B$ we find the mean-field equation
\be{meanf}
	B=H+\mvac(B)+\mmat(B)~~~.
\ee
This equation is telling us how to find the average microscopic field $B$, that
also is the acting field felt by the particles in the medium, in the presence
of
an external field $H$.

\Section{sec-vacuum}{The Vacuum of Spinor, Scalar and Vector QED}
We shall here consider the one-loop vacuum contribution to the
effective Lagrangian of spinor, scalar and vector
QED with an external static uniform  magnetic field ${\bf B }= B {\bf e}_z$.
In order to ease the physical interpretation, and the comparison with
\refc{nielsen81}, we shall use a cut-off regularization procedure throughout
this Section.

The Lagrangian for a spin 1/2 particle of charge $-e$ coupled to the
electro-magnetic field is
\be{lqed}
	\cL^{\rm ferm}= -\frac14 F_{\mu\nu}F^{\mu\nu} +
	\psibar (i D \slask -m) \psi~~~,
\ee
where the covariant derivative is $D_\nu = \del_\nu -i e A_\nu$, and we have
used the shorthand notation $D\slask\equiv \gamma^\mu D_\mu$. The
corresponding Lagrangian for the spin 0 case reads
\be{lscal}
	\cL^{\rm scal}= -\frac14 F_{\mu\nu}F^{\mu\nu} + D^*_\nu
	\phi^\dagger D^\nu \phi -m^2 \phi^\dagger \phi~~~.
\ee
For reasons that will become obvious in section~\ref{sec-qcd} we are also
interested in the case of a spin 1 vector particle with gyro-magnetic ratio
$\gamma=2$. The corresponding Lagrangian  is then obtained by adding
an anomalous magnetic moment term to the minimally coupled
Lagrangian~\cite{leeyang62}, with the result
\bea{lvec}
	\cL^{\rm vec}& =& -\frac14 F_{\mu\nu}F^{\mu\nu} - \frac12
	( D_\mu^* W_\nu^\dagger - D_\nu^* W_\mu^\dagger)
	( D^\mu W^\nu - D^\nu W^\mu) +
	m^2 W_\mu^\dagger W^\mu \\ \non
	&&  -ie (\gamma-1)F^{\mu\nu} \frac{W_\mu^\dagger
	W_\nu- W_\mu W_\nu^\dagger}2~~~.
\eea
The general quantized theory described by $\cL^{\rm vec}$ is not even
renormalizable~\cite{vanyashin65}.
But here we shall  consider no virtual photons. The renormalizability
then follows in exactly the same way as for scalar and spinor QED.
We separate the vector-potential into $A_\nu+\tilde A_\nu$, where $A_\nu$
is the vector-potential for ${\bf B}$, and $\tilde A_\nu$ corresponds to
the quantized radiation field.
We shall here only calculate the one-loop effective
Lagrangian. We may then neglect the radiation field, apart from the
kinetic term $-\tilde F^2/2$. For simplicity we shall not write out the
radiation field, but only add the contribution from thermal photons (or gluons)
in the
end.

A particle of
charge $-e$, spin {\boldmath $\sigma$} and gyro-magnetic ratio $\gamma$,
has the spin magnetic moment $ \mbox{\boldmath $\mu$}=-\gamma e \la
\mbox{\boldmath $\sigma$}\ra/2m $.
With spin projection $s$ along $-e {\bf B}$ the energy spectrum
reads~\cite{nknielseno78}
\be{genen}
	E_{n,s}=\sqrt{m^2+p_z^2+eB(2n+1-\gamma s)}~~~,
\ee
where the dependence on the momentum in the direction of the field ($p_z $)
has been suppressed, and we assume $eB>0$.

It is illustrated how  to  calculate the vacuum
contribution at one--loop level in spinor QED in
\refs{kaminski81,elmforsps93}, and also in scalar QED in \refc{elmforsps93}.
 In the general case we find similarly, using the trick to perform the
derivative
with respect to the mass
\be{genvac}
	 \frac{\del \cL_{\rm vac,b} }{\del m^2}= i(-1)^{2\sigma+1}
	\frac{eB}{(2\pi)^3}
	 \sum_{s}  \sum_{n=0}^\infty  \int
	d \om  d  p_z \frac{1}{w^2-E_{n,s}^2+ i\eps}~~~,
\ee
where $\sigma$ is the spin of the intermediate particles.
We have here suppressed that  the contribution at $B=0$ should be subtracted,
due to the normalization of the generating functional.
 The subscript $b$
denotes bare quantities, i.e. before renormalization.
The difference in sign comes from the fermionic Grassmann algebra.

 We may now use Cauchy's theorem to perform the integration
over $\om$. Integrating with respect to $m^2$ we find, with the subtraction
explicit
\be{vacenerg}
	 \cL_{\rm vac,b}= (-1)^{2\sigma+1} \frac{eB}{(2\pi)^2}
	 \sum_{s}  \sum_{n=0}^\infty  \int dp_z E_{n,s}-
	\cL_{\rm vac,b}(B=0)~~~.
\ee
In order to interpret this expression we shall calculate the density of states
for particles in a box $V=L_x L_y L_z$. We then need that the wave functions
in each of the theories here considered, with the choice of gauge
$A_\nu=(0,0,Bx,0)$,
 are of the generic form
\bea{wavefun}
  \Psi^{(\pm)}_{n,p_y,p_z,s}({\bf x},t)&=& \frac{1}{2\pi\sqrt{2E_{n,s}}}
        \, \exp[ \pm i(- E_{n,s} t \plus  p_{y}y \plus  p_{z}z ) ] \,
   I_{n;p_{y}}(x) ~~~, \\
	\label{Indef}
  I_{n;p_{y}}(x)& \equiv& \left( \frac{eB}{\pi} \right)^{1/4} \exp \left[
  - \frac{1}{2} eB \left( x \minus \frac{p_{y}}{eB} \right)^{2} \right]
 \frac{1}{ \sqrt{n!}} H_{n} \left[ \sqrt{2eB} \left( x \minus \frac{p_{y}}
 {eB} \right) \right]\; ,
\eea
where $H_n$ is a Hermite polynomial.
 Without the magnetic field, the number
 of states on the momentum interval $dp_i$ is $L_i/2\pi\, dp_i$. With the above
magnetic field, this still holds true for $i=z$.
 The wave-functions in \eq{Indef}
corresponds to a harmonic oscillator, centered at $x_0=p_y/eB$, that has to
be within the box, i.e. $0\leq p_y/eB \leq L_x$. Since the energy is
independent of $p_y$ we must sum over all possible values. The resulting
degeneracy of states is then
\be{degen}
	V \frac{eB}{(2\pi)^2}~~~.
\ee
The vacuum Lagrangian density is thus equal to the negative vacuum energy
density
\be{vacen}
	 \cL_{\rm vac,b}= \frac{-1}V [ E_{\rm vac} - E_{\rm vac}(B=0)]~~~,
\ee
again without the contribution at vanishing magnetic field.
The vacuum energy was the starting point in \refc{nielsen81}. The vacuum
energy  needs to be regularized. As it stands it is quadratically divergent,
but when the $B=0$ part has been subtracted the result is only logarithmically
divergent.

Instead of immediately integrating with respect
to $m^2$, we shall here first use
\be{sqrtid}
	\frac1{2E} = \frac{1}{\sqrt{\pi}} \int_{1/\Lambda}^\infty \,
	 \exp[-E^2 x^2]
	dx + \cO(\frac1\Lambda)~~~,
\ee
where we have introduced the ultra-violet cut-off $\Lambda$, that essentially
removes the contributions for $E > \Lambda$ due to the exponential suppression.
We may now perform the Gaussian integral over $p_z$, sum
the infinite geometrical
series in $n$, and integrate with respect to $m^2$. Subtracting the
contribution for $B=0$, and changing variable of integration to $t=x^2$, we
arrive at
\be{lvacgenfin}
   \cL_{\rm vac,b}=\frac{(-1)^{2\sigma}}{16\pi^2} \sum_s
	\int_{1/\Lambda^2}^\infty
	\frac{dt}{t^3} \exp(-m^2 t) \left\{ \frac{eBt}{\sinh(eBt)}
	\exp(eBt \gamma s )-1 \right\}~~~.
\ee
For $eB(\gamma s - 1)>m^2 $ this is divergent
for large $t$. This is  the case for large fields
 in vector QED ($\sigma=1$) with
$\gamma=2$, that will be treated separately below. Let us now first consider
the massive case in Section~\ref{sec-massvac}, and then the mass-less case in
Section~\ref{sec-lessvac}.\\

\Subsection{sec-massvac}{The Massive Case}
In the limit as $x\equiv eBt \rightarrow 0$, we have
\be{lim}
	\frac{x}{\sinh(x)} \exp(\gamma s x)=1 + \gamma s x +\frac12 x^2
	\left(\gamma^2 s^2-\frac13 \right) + \cO(x^3)~~~.
\ee
Notice that $\sum_s s=0$. Define from the quadratic term
\be{xidef}
	\ch \equiv \frac{(-1)^{2\sigma}}{16\pi^2} \sum_s \left(\gamma^2 s^2 -
	\frac13 \right)~~~.
\ee
Let us now subtract and add the term containing
\be{sub}
	\int_{1/\Lambda^2}^\infty \frac{dt}{t} \exp(-m^2 t) =E_1(m^2/\Lambda^2)
	= -\gamma_E + \ln\left( \frac{\Lambda^2}{m^2} \right) +\cO\left(
	\frac{m^2}{\Lambda^2} \right)~~~,
\ee
where $E_1$ is an exponential integral, and $\gamma_E =
0.57721566 \ldots$ is Euler's constant.
In order to get rid of the cutoff $\Lambda$ we must now perform a
renormalization. The Ward identities of the theories here considered requires
the product $eB$ to be invariant. Let us therefore
rescale the external field and the coupling according to
\bea{renorm}
	e_b^2&=&Z_\lambda^{-1} e^2(\lambda)~~~, \\
	B_b&=&Z_{\lambda}^{1/2} B(\lambda)~~~,
\eea
where $e(\lambda)$ and $B(\lambda)$ are the charge and background field
renormalized at momentum scale $\lambda$, respectively.
 Adding
$0=\ln(\lambda^2/\lambda^2)$, we may write
\be{leffag}
	\leff=-\frac12 \frac{(eB)^2}{e^2(\lambda)} -\frac12(Z_{\lambda}-1)
	\frac{(eB)^2}{e^2(\lambda)}
	+\frac12 (eB)^2 \ch \left[ -\gamma_E +
	\ln \!\left(\frac{\Lambda^2}{\lambda^2}\right) \right]+
	\lvac(eB,\lam)~~~,
\ee
where the finite, renormalized vacuum contribution is
\bea{renvac}
	\lvac(eB,\lam)& =& \frac12 (eB)^2 \ch\ln \!\left(
	\frac{\lambda^2}{m^2}\right) +\tilde \lvac(eB)~~~,\\
	\tilde \lvac(eB)&=& \int_{0}^\infty
	\frac{dt}{t^3} \exp(-m^2 t) \left\{
	\frac{(-1)^{2\sigma}}{16\pi^2} \sum_s \left[ \exp(eBt \gamma s )
	 \frac{eBt}{\sinh(eBt)}-1
	\right] - \ch \frac{(eBt)^2}{2} \right\}~. \non \\
	&&
\label{tildevac}
\eea
We must now choose $Z_\lambda$ in such a way that the divergence as
$\Lambda \rightarrow \infty$ is removed
\be{Zchoice}
  	Z_{\lambda}-1=e^2(\lambda^2)\ch \left[ -\gamma_E +
	\ln \!\left(\frac{\Lambda^2}{\lambda^2}\right) \right]~~~,
\ee
i.e.
\be{chargeren}
	\frac{1}{e^2(\lambda)}= \frac{1}{e_{b}^2} +
	\ch \left[ -\gamma_E -
	\ln \!\left(\frac{\Lambda^2}{\lambda^2}\right) \right]~~~.
\ee
Performing the derivative with respect to $\lambda$ we find the  lowest
order  $\beta$-function
\be{betafun}
  	\lambda\frac{de(\lambda)}{d \lambda} \equiv
	\beta[e(\lambda)]=-\ch e^3(\lambda)~~~.
\ee
In the case of (spinor) QED, $\sigma=1/2$ and $\gamma=2$, we find
\be{cqed}
	\ch^{\rm ferm}=-\frac{1}{12\pi^2}~~~.
\ee
Similarly in scalar QED, $\sigma=0$ gives
\be{scalqed}
	\ch^{\rm scal}=-\frac{1}{48\pi^2}~~~.
\ee
Inserting this into \eq{betafun}, we recognize the correct $\beta$-function
of spinor and scalar QED. In the theory of vector QED
with gyro-magnetic ratio $\gamma=2$, the $\beta$-function really does not
exist,
due to the lack of renormalizability. However,  for vector bosons interacting
only
with the external field we find
\be{xivacmass}
	\left. \ch^{\rm vec}\right|_{\rm massive}=\frac{7}{16\pi^2}~~~.
\ee
This does not agree  with the renormalization of the
coupling  considered in \refc{vanyashin65} (corresponding to $ \ch^{\rm vec}=
5/16\pi^2$). Notice the crucial difference in sign of $\ch$ for vector QED,
that we
later shall relate to QCD.
The solution  to the above renormalization group equation may be written
\be{rensol}
	\frac1{e^2(\lambda)} = \frac1{e^2(\lambda_0)} + \ch
	\ln \left( \frac{\lambda^2}{\lambda_0^2} \right)~~~.
\ee
With $\ch<0$  this corresponds to the effective charge {\em increasing}
 with the momentum scale $\lam$, whereas  $\ch>0$ corresponds to
the  effective charge {\em decreasing}.
Obviously $e(\lambda)$ and $B(\lambda)$ are defined such that $\leff$ is
independent of $\lambda$. We  have
\be{fingenleff}
	\leff=-\frac12 \frac{(eB)^2}{e^2(\lambda)} +\frac12 (eB)^2
	\ch \, \ln \!\left( \frac{\lambda^2}{m^2} \right)
	+\tilde\lvac(eB)~~~.
\ee
Notice that $Z_\lambda$ only depends on $\lambda$ through the
 dimension-less
parameter $\lambda/\Lambda$. As long as the $\Lambda$ dependence is
absorbed in $Z_\lambda$ (as is necessary to get finite expressions) the
RGE(\ref{betafun}) will not depend on how $Z_\lambda$ and thus $e(\lambda)$
are defined.
Since the cutoff $\Lambda$ no longer appears in the effective Lagrangian,
we may now send it back to infinity where it belongs.

In order to obtain a physically clear picture let us now consider the high
field limit.
Substitute $y=eBt$ in \eq{tildevac},
and split the integral at $y_0$, such that $eB/m^2 \gg y_0 \gg 1$.
The contribution from
$y<y_0$ is easily seen to be $\cO[(eB)^2]$. The contribution from $y>y_0$ is
dominated by the subtracted
term $\ch  (eBt)^2/2$, that gives
\be{highbgen}
	\frac{\tilde\lvac}{(eB)^2} \simeq -\frac12  \ch \,
	\frac{m^2}{(eB)^2}\, \int_{y_0}^{\infty} \frac{dy}y \exp\left(
	-\frac{m^2}{eB}y \right) \simeq -\frac12
	\ch \, \ln\left( \frac{eB}{m^2}
	\right)~~~.
\ee
At the energy scale $eB=\lambda^2\gg m^2$, the two logarithms will cancel,
and thus the effective Lagrangian assumes a purely Maxwellian form
\be{freeleff}
	 \leff \simeq
	-\frac12 B^2(\lambda^2=eB)~~~,~eB \gg m^2~~~.
\ee
We thus find that at least in the limit $\lambda \gg m$, that $e(\lambda)$ and
$B(\lambda)$ are the charge and field strength measured at momentum scale
$eB=\lambda^2$, respectively.

In the high field limit ($eB\gg m^2$), \eq{highbgen} gives to leading order
\bea{highcm}
	\frac{\mvac(\lam)}{e(\lambda)} &= &- eB \, \ch \,
	 \ln \!\left(\frac{eB}{\lambda^2}\right)~~~,\\
	\cvac(\lam) &=& - e^2(\lam)
	\, \ch \, \ln \!\left(\frac{eB}{\lam^2}\right)~~~.
\eea
We see that the sign of the magnetic susceptibility in addition to $\ch$
depends on the relative
magnitude between $eB$ and $\lam^2$.
 This does anyhow correspond to screening in spinor and scalar QED ($\ch<0$);
 but anti-screening in vector QED ($\ch>0$)  since
\be{chider}
	\frac{\;\;\partial \cvac(\lam)}{\partial (eB)}=- e^2(\lam)
	 \ch \, \frac1{eB} ~~~.
\ee
The permittivity $\varepsilon$ is thus  decreasing with the energy scale $eB$
in spinor and scalar QED (with $\ch<0$) , but increasing in vector QED (with
$\ch>0$).
\smallskip

 In the high field limit we find the effective charge
\be{aeffhigh}
	\frac1{\eeff^2(eB)} \simeq \frac1{e^2(\lam_0)}+\ch
		\ln \!\left(\frac{eB}{\lam_0^2}\right)~~~,~eB\gg m^2~~~,
\ee
that is a solution to the  lowest order RGE with the momentum scale
identified as $\lambda=\sqrt{eB}$. Notice that $\eeff$ must be independent
of $\lam_0$ by its definition in \eq{aeffdef}, since $\leff$ and $eB$ so is, and
that
this follows from \eq{rensol}.
 It is no coincidence that the above effective coupling
satisfies the lowest order RGE, since it corresponds to a summation of the
leading logarithms. The high field limit is dominated by the  term containing
$\ch$,
subtracted in the renormalization, that is defining the
$\beta$--function.\bigskip

The effective Lagrangian will assume a particularly simple
form  if we choose $\lambda=m$
\be{leffsimple}
	\leff =-\frac12 B^2 +\tilde\lvac(eB)~~~.
\ee
Here we have suppressed the dependence on  $\lambda$, since in the weak field
limit
\be{leffweak}
	\leff=-\frac12 B^2 + B^2 \cO\left[e^2 \frac{(eB)^2 }{m^4}\right]~~~,
\ee
so that $e$ and $B$ are the ordinary electric charge and magnetic field,
measured at vanishingly small magnetic field, corresponding to long
wavelengths. The  vacuum of spinor and scalar QED
 will thus be paramagnetic ($\cvac>0$) in the high field limit $\{eB \gg
\lam^2=m^2\}$.
Actually, in the spinor case it is paramagnetic for all values of the magnetic
field, as shown in
\refc{elmforslps95}, to which we refer for a more extensive treatment of
the vacuum magnetization in spinor and scalar QED.
In spinor QED we have the vacuum contribution to the effective Lagrangian
 explicitly
\be{qedvac}
	\lvacq= -\frac1{8\pi^2} \int_0^\infty \frac{dt}{t^3} \exp(-m^2t)
	\left[ eBt \coth(eBt)-1-\frac13 (eBt)^2 \right]~~~.
\ee
In scalar QED we find
\be{scalvac}
	\lvacs= \frac1{16\pi^2} \int_0^\infty \frac{dt}{t^3} \exp(-m^2t)
	\left[ \frac{eBt}{ \sinh(eBt)}-1+\frac16 (eBt)^2 \right]~~~.
\ee
\bigskip

As remarked above the case of vector QED for $eB>m^2$ needs more careful
considerations.
Let us use the similarity between vector and scalar QED to write
\be{barevec}
	\left. \cL_{\rm vac,b}^{\rm vec}\right|_{\rm massive}=
	3 \cL_{\rm vac,b}^{\rm scal} +
	\Delta \cL_{\rm vac,b }^{\rm vec} ~~~.
\ee
Written on the vacuum energy form corresponding to \eq{vacenerg} the extra
term reads
\be{deltabare}
	\Delta \cL_{\rm vac,b }^{\rm vec}= -\frac{eB}{(2\pi)^2}
	\int dp_z (\sqrt{m^2+p_z^2-eB-i \eps}-\sqrt{m^2+p_z^2+eB}~)~~~.
\ee
Obviously, this will contain an imaginary part for $eB>m^2$
\be{impart}
	\Im \dlvac = \frac{eB}{2\pi^2} \int_{-\sqrt{eB-m^2}}^{\sqrt{eB-m^2}}
	dp_z \sqrt{eB-m^2-p_z^2}=
	\Theta(eB-m^2) \frac{(eB-m^2)eB}{8\pi}~~~.
\ee
Working out the real part in analogy to the general case we find
\bea{deltavec}
	\Delta \cL_{\rm vac,b }^{\rm vec}&=& \frac12 \frac{(eB)^2}{2\pi^2}
	\ln \left( \frac{\Lambda^2}{\sqrt{(m^2+eB)|m^2-eB|}} \right) -
	\frac{eBm^2}{8\pi^2}\left[ \ln\left( \frac{m^2+eB}{|m^2-eB|}\right) -
	2 \frac{eB}{m^2}\right] \non \\
	&& - C (eB)^2 + i
	\Theta(eB-m^2) \frac{(eB-m^2)eB}{8\pi}~~~,
\eea
where $C$ is an irrelevant constant that anyhow will be renormalized away.
Notice that the result also may be obtained by analytical continuation
from the result valid for $m^2>eB$, i.e. by substituting $|m^2-eB| \mapsto
	(m^2-eB-i\eps)$.
Also notice that the dependence on the cut-off $\Lambda$ is independent of the
 magnitude of $m^2-eB$, so the renormalization will not be affected.
The imaginary part in the effective Lagrangian is telling us that the
configuration of a spin 1, gyro-magnetic ratio 2 particle in a strong
$(eB>m^2$) static
uniform magnetic field is unstable. This problem has been addressed earlier
in the literature, see Section~\ref{sec-qcd} for references and
a discussion. The conclusion is that
a condensate will be formed in the unstable mode. The magnetization
of this condensate will alter the background magnetic field so that it is
no longer uniform on a microscopic scale. The energy of the lowest mode
will then become positive so that  the instability is removed. However, the net
change in the magnetic field is found to be small, so we simply neglect
this here.  We are thus assuming our results to be valid also in vector QED,
and neglect the appearance of the imaginary part. The coefficient $\ch$
in front of $(eB)^2\ln(\Lambda^2/m^2)$ obtained from \eq{barevec} agrees with
\eq{xivacmass}.

\Subsection{sec-lessvac}{The Mass-less Case}
Let us now consider the mass-less limit of spinor, scalar and vector QED.
For vector bosons it is then necessary to choose the gauge $D^\mu W_\mu=0$,
in order for the wave-functions to be of the same  form as in \eq{wavefun}.
This will actually introduce Faddeev--Popov ghosts. In for example
\refc{ninomiyas81} it was shown that the contribution from these ghosts
exactly cancel the contribution from the unphysical degrees of freedom.
This means that if we restrict the sum over polarizations
 to $s=\pm 1$, we do not need to consider ghosts.

In this mass-less
case we may not subtract the next term in the expansion of \eq{lim},
since this would cause a divergence for large $t$. Instead  we substitute
$x=eBt$ in \eq{lvacgenfin} and  integrate by
parts to find
\be{leffless}
	\cL_{\rm vac,b}= \frac12 \ch (eB)^2 \ln \left( \frac{\Lambda^2}{eB}
	\right) + \frac{C'}2 (eB)^2~~~,
\ee
where the u.v. finite constant is
\be{cdef}
	C' \equiv \frac{(-1)^{2\sigma}}{8\pi^2} \sum_s \int_0^\infty dx
	\ln x\, \frac{d}{dx} \left[ \frac{ \exp(\gamma s x) }{x\sinh x}
	-\frac1{x^2}  \right]~~~.
\ee
In the mass-less limit, vector QED will be unstable for all values of $B$.
Since  only  two different polarizations $s=\pm 1$ now are possible,
we write
\be{vecless}
	\lvacv=2\lvacs + \dlvac~~~.
\ee
We may directly take the mass-less limit in \eq{deltavec}. The result is
that \eq{leffless} still is valid, but with an imaginary part in the
constant $C'$.
This is of no surprise since the renormalization is due to ultra-violet
divergences, i.e. $p_z^2 \gg eB$.
 Again we shall neglect this imaginary part and the
instability. The effective Lagrangian of \eq{leffless} is what was considered
in \refc{nielsen81} (but in terms of the vacuum energy).
Nielsen~\cite{nielsen81} extracted from here the leading bare vacuum magnetic
 susceptibility
\be{baresusc}
	\chi_{\rm vac,b} \simeq  e_b^2 \ch \ln \frac{\Lambda^2}{eB}~~~.
\ee
For mass-less vector bosons we find explicitly
\be{chivec}
	\ch^{\rm vec} = \frac{11}{24\pi^2}~~~.
\ee
This has the opposite sign as  compared to spinor and scalar QED.
Bare vector QED (that we in  Section~\ref{sec-qcd} will  relate to QCD)
will thus be paramagnetic, that corresponds to anti-screening and
asymptotic freedom. However, performing the renormalization we have
\be{renless}
	\lvac=-\frac12 \ch (eB)^2 \ln \frac{eB}{\lam^2}~~~.
\ee
The sign of the  renormalized vacuum magnetic susceptibility, i.e. if we have a
paramagnetic ($\cvac>0$) or diamagnetic ($\cvac<0$) vacuum,  will
thus again depend on the relative magnitude between the field strength $eB$,
and the renormalization scale $\lam^2$.\bigskip

Asymptotic freedom really means anti-screening with the effective charge
vanishing at vanishing distance (infinitely large momentum scale). From the
RGE this is obvious since $e(\lam)=0$ is a fixed point. We may also compare
the effective charge at two different scales
\be{effcomp}
	\frac{\eeff^2(eB)}{\eeff^2(eB_0)}=
	\left[ 1+\eeff^2(eB_0) \ch \ln \! \left(\frac{eB}{eB_0}
		 \right) \right]^{-1}~~~.
\ee
If $\ch >0$, as in vector QED and QCD this quotient  is vanishing as
 $eB \rightarrow \infty$,
i.e. we have asymptotic freedom. In spinor and scalar
QED with $\ch <0$, the coupling appears
to  grow
infinitely large at a finite (but extraordinary large) value of the magnetic
field. However, perturbation theory (we are only considering one loop effects
here) cannot be extrapolated that far.

\Section{sec-thermal}{Thermal  Spinor, Scalar and Vector QED }
We shall in this section consider the contributions to the effective
Lagrangian due to finite temperature and density of particles.
 We may
on this one loop level use the naive real time formalism. In \refc{elmforsps93}
it was shown that the substitution in the fermion vacuum propagator
\be{fermsubst}
	f_F(\om)=
	\frac{i}{\om^2-E^2+i\eps} \mapsto- 2\pi  \delta(\om^2-E^2) f_F(
	\om)~~~,
\ee
in order to obtain the thermal part of the propagator, works also in a
magnetic field. Here the fermion  one-particle  distribution
in thermal equilibrium is the Fermi--Dirac distribution
\be{fermidirac}
	f_F(\om)= \frac{\Theta(\om)}{e^{\beta(\om-\mu)}+1}
	+ \frac{\Theta(-\om)}{e^{\beta(-\om+\mu)}+1}~~~,
\ee
where $\mu$ is the chemical potential, and $1/\beta=T$ is the temperature.
In the bosonic case we substitute similarly
\be{bosesubst}
	\frac{i}{\om^2-E^2+i\eps} \mapsto 2\pi  \delta(\om^2-E^2) f_B(
	\om)~~~,
\ee
where $ f_B$, in thermal equilibrium, is the Bose--Einstein distribution
\be{boseeinstein}
f_B(\om)= \frac{\Theta(\om)}{e^{\beta(\om-\mu)}-1}
	+ \frac{\Theta(-\om)}{e^{\beta(-\om+\mu)}-1}~~~.
\ee
Performing these substitutions in \eq{genvac}, we may use the $\delta$-function
to integrate over $\om$. Integrating with respect to $m^2$, the constant of
 integration is determined by the fact that the exponential suppression from
the particle distributions requires $\lmat \rightarrow 0$, as
$m \rightarrow \infty$. We then find the contribution from the heat and
charge bath to the effective Lagrangian
\be{genmat}
	\lmat=\frac{eB}{(2\pi)^2} \sum_s \sum_{n=0}^\infty \int dp_z
	\frac{p_z^2}{E_{n,s}} \left[ \frac1{e^{\beta(E_{n,s}-\mu)} -
	(-1)^{2\sigma}} + \frac1{e^{\beta(E_{n,s}+\mu)} -
	(-1)^{2\sigma}}\right]~~~.
\ee
Integrating by parts with respect to $p_z$ we recognize the free-energy density
\bea{genfreen}
	\lmat \equiv  \frac{1}{\beta V} \ln Z&=&
	\frac{(-1)^{2\sigma+1}}{\beta}
	\frac{eB}{(2\pi)^2} \sum_s \sum_{n=0}^\infty \int dp_z
	\left\{ \ln \left[ 1- (-1)^{2\sigma}e^{-\beta(E_{n,s}-\mu)}
	\right] \right. \non \\
	&&\left.+\ln \left[ 1- (-1)^{2\sigma}
	e^{-\beta(E_{n,s}+\mu)}
	\right]\right\}~~~,
\eea
where $Z$ is the partition function of the gas. Let us now split $\lmat$ into
\be{lmatsplit}
	\lmat\equiv \lmatn +\lmate~~~,
\ee
where $\lmatn$ is the field independent part
\be{lmat0}
	\lmatn=\frac{(-1)^{2\sigma+1}}{\beta} \sum_s \int \frac{d^3 {\bf p}}{
	(2\pi)^3}
	\left\{ \ln \left[ 1- (-1)^{2\sigma}e^{-\beta(E(p)-\mu)} \right]+
 	\ln \left[ 1- (-1)^{2\sigma}e^{-\beta(E(p)+\mu)} \right]
	\right\}~,
\ee
whith $E(p)=\sqrt{m^2+{\bf p}^2}$.
We are particularly interested in the high temperature expansion of the
effective Lagrangian.
 We shall consider the massive
 case for spinor, scalar and vector QED in Section~\ref{sec-massive}, and
 investigate
 the corresponding mass-less
limits in  Section~\ref{sec-massless}.
In every case the contribution from  thermal photons
\be{termph}
	\cL_{\rm mat}^{\rm phot}=\frac{2\pi^2}{45}T^4~~~,
\ee
should be added to the effective Lagrangian.
In Section~\ref{sec-dense} we consider also a degenerate fermion gas at low
temperature
and high density.\\

\Subsection{sec-massive}{The Massive case}
The case of massive fermions in a magnetic field was treated in
\refc{elmforsps93}. Here we just quote the final result
\bea{termferm}
	\lmate^{\rm ferm}&=& \int d\om \Theta(\om^2-m^2) f_F(\om)
	\left\{ \frac1{4\pi^{5/2}} \int_0^\infty \frac{dt}{t^{5/2}} e^{-t(
	\om^2-m^2)}[eBt \coth(eBt)-1] \right. \non \\
	&& \left. -\frac1{2\pi^3} \sum_{l=1}^\infty \left( \frac{eB}l
	\right)^{3/2} \sin\left( \frac\pi4-\pi l \frac{\om^2-m^2}{eB} \right)
	\right\}~~~.
\eea

The high temperature, weak field limit
$\{T^2 \gg m^2 \gg eB;~ \mu=0\}$  was considered in
\refc{elmforslps95}, with the result
\be{masshight}
	\lmatq +\tilde\lvacq  = \frac{7\pi^2}{180} T^4 -
	\frac12 \ch^{\rm ferm}(eB)^2
	\ln \left( \frac{T^2}{m^2} \right) + \cO[(eB)^2]~~~.
\ee
Notice here that in  the high-temperature effective Lagrangian,
terms with higher powers of $eB$ are  suppressed as $ (eB)^2
(eB/T^2)^{n-2},~~~n \geq 4$. To leading order there are thus no non-linear
electro-magnetic interactions in high temperature QED, in agreement
with a diagrammatic analysis~\cite{brandtf95}. Also notice that
the $\ln m^2$ terms will cancel when this is added to the effective Lagrangian
in \eq{fingenleff}, and that
the effective charge obtained from this will satisfy the lowest order RGE
with the scale $\lambda=T$.
\bigskip

Let us now consider the case of massive  scalar QED. For the sake of
completeness we here quote the result (see e.g. \refc{elmforsps93})
\bea{termbose}
	\lmate^{\rm scal}&=& \int d\om \Theta(\om^2-m^2-eB) f_B(\om)
	\left\{ \frac1{8\pi^{5/2}} \int_0^\infty \frac{dt}{t^{5/2}} e^{-t(
	\om^2-m^2)}\left[\frac{eBt}{ \sinh(eBt)}-1\right] \right. \non \\
	&& \left. -\frac1{4\pi^3} \sum_{l=1}^\infty \left( \frac{eB}l
	\right)^{3/2} \sin\left( \frac\pi4-\pi l \frac{\om^2-m^2-eB}{eB}
	\right)	\right\}~~~.
\eea
In the  high temperature, weak field  limit,  this gives~\cite{elmforslps95}
\bea{bosehight}
	 \lmats+ \tilde\lvacs&=& \frac{\pi^2}{45} T^4 -
	  \frac12 \ch (eB)^2  \left\{ \ln \left(\frac{4\pi T}{m}
	\right)^2 -\gamma_E +\cO\left[ \left(\frac{m}T \right)^2 \right]
	\right\} \non \\
	&& -\frac{T}{m}  \frac{(eB)^2}{48\pi} \left\{ 1+ \cO \left[ \left(
	\frac{eB}{m^2} \right)^2 \right] + \cO\left[ \frac{m}T \left( \frac{
	eB}{T^2} \right)^2 \right]\right\}~~~.
\eea
In the massive vector case we again write $\lmatv=3\lmats + \dlmat$.
For  $m^2>eB$ and $\mu=0$, we have
\be{deltmass}
	\dlmat=\frac{eB}{\pi^2} \int_0^\infty dp_z p_z^2 \left[
	 \frac1{E_{0,-1}} \frac1{e^{\beta E_{0,-1}}-1} - \frac1{E_{0,0}}
	\frac1{e^{\beta E_{0,0}}-1} \right]~~~.
\ee
The calculations performed in order to find the high temperature expansion of
this
expression are explicitly shown in Appendix~\ref{app-deltmass}. The final
result
reads
\bea{dmassfin}
	\dlmat& =& -\frac{(eB)^2}{4\pi^2} \ln \left[ \frac{T^2(4\pi)^2}{\sqrt{
	|m^4-(eB)^2|}}\right] +\frac{eBT}{2\pi} (\sqrt{m^2+eB}- \sqrt{|m^2-eB|} )
	 \non \\
	&& + \frac{eB m^2}{8\pi^2} \ln \left( \frac{m^2+eB}{|m^2-eB|} \right)
	-\frac{(eB)^2}{2\pi^2} \left( \frac12 -\gamma_E \right)
	+eBT^2 \cO\left(\frac{m^2+eB}{T^2} \right)^{3/2}\non \\
	&& +i \Theta(eB-m^2)\left[\frac{eB T\sqrt{eB-m^2}}{2\pi} -
	\frac{eB(eB-m^2)}{8\pi}
	\right]~~~.
\eea
Notice that the $\ln(m^2\pm eB)$ terms  cancel between \eq{deltavec}
and \eq{dmassfin}.
For $eB>m^2$ the lowest energy mode will become unstable, resulting in the
imaginary
part as depicted above. The result follows also in this case by analytical
continuation
$m^2 \rightarrow m^2-i\eps$ in expression valid for $eB<m^2$.
Notice that the $T$-independent imaginary part exactly cancels
the imaginary part of the corresponding vacuum contribution in \eq{deltavec},
and
that the resulting imaginary part is increasing linearly with the temperature.

Of course such an imaginary part is not acceptable. Since a full treatment of
this
imaginary part is out of the scope here we are left with two alternatives.
Either neglecting  the imaginary part, or neglecting the total contribution
from
the tachyonic mode, that also has a real part. In  the former approach we may
immediately use \eq{dmassfin}. The latter approach is considered in the
mass-less
 case in
the following Section~\ref{sec-massless}. The massive case (that is relevant
for
spontaneously broken gauge theories) can be treated in full analogy.

\Subsection{sec-massless}{The Mass-less case}
The explicit calculations performed in order to obtain the high temperature
expansion
in mass-less spinor and scalar QED are presented in Appendix~\ref{app-less}.
 Adding the thermal contribution for $B=0$,
$\lmatn$, The final result reads in the spinor case
\be{lessspin}
	\lvacq+\lmatq=\frac{7\pi^2}{180}T^4 -\frac12 \ch^{\rm ferm} (eB)^2 \ln
	\frac{T^2}{\lambda^2} +\cO[(eB)^2]~~~.
\ee
Just like in the massive case, all non-linear electro-magnetic
interactions from the vacuum part has been cancelled by contributions from
the thermal part.
In the scalar case we find
\be{lessscal}
	\lvacs+\lmats=\frac{\pi^2}{45}T^4
	 -\frac{\sqrt2 -1}{2\pi} |\zeta(-\frac12)|
	(eB)^{3/2}T -\frac12 \ch^{\rm scal} (eB)^2 \ln
	\frac{T^2}{\lambda^2}  +\cO[(eB)^2]~~~.
\ee
Again the $\ln[(eB)^2/\lambda^2]$ has been cancelled by thermal contributions,
but there is in this case also the term linear in $T$.\bigskip

We are now left with the vector bosons.
The configuration of mass-less vector bosons in a static uniform magnetic field
is
unstable for all field-strengths.
 We will here  neglect
the contribution for $p_z^2<eB-m^2$ in the lowest Landau level in order to
avoid the imaginary part. However,
this  may affect the high temperature behavior, that is not solely
governed by large
momenta, but also by small energies when the Bose-Einstein distribution is
becoming very large. The final result of the calculations performed in
Appendix~\ref{app-deltless} reads
\be{deltlessfin}
	\dlmat= -\frac{(eB)^2}{\pi^2} \left\{ \frac{T}{\sqrt{eB}} \left[
	\frac12 \ln \frac{T^2}{4eB} +1-\frac\pi2 \right] +
	\frac14 \ln \frac{T^2}{eB}
	+\cO(1) \right\}~~~.
\ee
The same expression is obtained if we substitute $m^2 \rightarrow -i \eps$
in \eq{dmassfin}, and subtract the corresponding integral $\int_0^{\sqrt{eB}}$
analytically continued to imaginary energy.

\Subsection{sec-dense}{The Dense Degenerate fermion Gas}
Chodos et al.~\cite{chodosko94} has suggested that the fermion matter
contribution in
the limit of large chemical potential  may be of importance in
heavy-ion collisions. Let for simplicity $T=0$. The leading behavior
of the first term in \eq{termferm} is then~\cite{elmforsps93}
\be{highmu}
	\cL_{\rm reg}\simeq -\frac12 \ch^{\rm ferm}(eB)^2 \ln
	\frac{\mu^2}{m^2}~~~.
\ee
We may  write the
oscillating second term in \eq{termferm} as
\be{osc}
	\cL_{\rm osc} = -\frac{(eB)^{3/2}}{2\pi^3} \sum_{n=1}^\infty
	\frac1{n^{3/2}} \, \Im \left\{ \exp\left[i\left( \frac\pi4-
	\pi n \frac{m^2}{eB} \right) \right]  J_n \right\}~~~.
\ee
Here we have defined
\bea{jndef}
	J_n& \equiv & \int_m^\mu d\om \exp\left[ -i \pi n \frac{\om^2}{eB}
	\right] \non \\
	&=& e^{-i\pi/4} \sqrt{\frac{eB}{2n}}\left\{ \erf\left[e^{i\pi/4}
	\sqrt{\frac{n\pi \mu^2}{eB}} \right] - \erf\left[e^{i\pi/4}
	\sqrt{\frac{n\pi m^2}{eB}} \right] \right\}~~~,
\eea
where error functions were identified. Let us now neglect the second term
that is independent of the chemical potential. Using the asymptotic expansion
of the error function~\cite{hmf}
 for $\pi \mu^2 \gg eB$ we find the leading term at
large chemical potentials
\be{loscmu}
	\cL_{\rm osc} \simeq - \frac{(eB)^{5/2}}{4\pi^4\mu } \sum_{n=1}^\infty
	\frac1{n^{5/2}} \sin\left( \frac\pi{4} + n \pi
	\frac{\mu^2-m^2}{eB} \right)~~~.
\ee
This is suppressed at large $\mu$, but performing the derivative with respect
 to $eB$ we find the leading behavior, with ${\rm mod}[A] \equiv A -
{\rm int}[A]$,
\bea{moscmu}
	\frac{ M_{\rm osc}}{e^2B} & \simeq& \frac{\mu}{\sqrt{eB} }
	\frac{1}{4 \pi^3}  \sum_{n=1}^\infty
	\frac1{n^{3/2}} \sin\left( \frac\pi{4} -2 n \pi
	\frac{\mu^2-m^2}{2eB} \right) \\
	&= & -\frac{\mu }{\sqrt{2eB}}\frac1{3\pi^2} \zeta\left( -\frac12, {\rm mod}
	\left[\frac{\mu^2-m^2}{2eB}\right] \right).
\eea
In \refc{chodosko94} the results of \refc{elmforsps93}
was used to find  \eq{moscmu}, using another method. Chodos et
al.~\cite{chodosko94} suggest that the effective charge obtained from this,
increasing linearly with $\mu$ for $\mu^2=m^2+2eBn,~n\in {\bf Z}$,
could be of relevance in heavy ion collisions. In \refc{chodosko94} it was
claimed that this is valid only for $\mu^2\gg eB \gg m^2$, but here we have
shown that this is the leading behavior for $\mu^2\gg eB,~m^2$
irrespective of the relative  magnitude of  $eB$ and $m^2$.
The effective charge then seems to become divergent as $eB \rightarrow 0$.
This must be an artifact due to one out  of two possibilities.
\begin{enumerate}
\item The break-down of perturbation theory, as the coupling is becoming
	 stronger.

\item The derivative in the definition of the effective charge. If we instead
would define the effective charge by $1/e_{\rm eff}^2= -2 \leff/(eB)^2$,
then the contribution from the oscillating part is suppressed at large
chemical potentials.
\end{enumerate}
Furthermore, the linearly increasing amplitude in $M_{\rm osc}$ is a result of
the sharp de Haas--van Alphen oscillations at the Fermi surface, that
will be smoothed out
 at finite temperature.
The $\zeta$-function in \eq{moscmu} takes its
minimal value $\zeta(-1/2,0)=\zeta(-1/2,1) \simeq
-0.208$, and its maximal value $\zeta(-1/2,0.3027) \simeq 0.0934$.
The effective charge \bort{obtained in this limit may thus be smaller or larger
than the reference charge, depending on the value of the quotient
$(\mu^2-m^2)/(2eB)$} is thus oscillating,
 and we doubt its physical significance in this limit.

\Section{sec-qcd}{QCD with a  Background Magnetic Field}
Quantum Chromodynamics with a background magnetic field has been extensively
studied in the literature.
It is well-known that the mean-field equation~(\ref{meanf})  has a
non-vanishing solution for $B$, even in the absence of an external field
(i.e.  $H=0$) in QCD.
The vacuum energy is lower in this state than for $B=0$, as pointed out
in \refc{savvidy77}. However a tachyonic mode appears , cf. \eq{genen},
that causes an imaginary part in the effective potential~\cite{nknielseno78},
signaling the instability of this configuration. This unstable mode has been
suggested to be removed by a (1+1) dimensional dynamical Higgs
mechanism~\cite{ambjornno79}. The corresponding condensate shows a domain-like
structure, and may form a quantum liquid of magnetic flux
tubes~\cite{hbnielseno79,ambjorno80}. A corresponding treatment has also been
done in the electro-weak theory (e.g. \refs{ambjorno90,macdowellt92}). Here
the change in the magnetic field due to the condensate, that assumes a
lattice structure, is found to be small compared to the uniform field.
It has also been suggested in another approach that the tachyonic mode could be
stabilized by radiative corrections~\cite{skalozub80,reznikov83}, or
by the condensation of an auxiliary field~\cite{ghoroku80}. In
 \refs{kaisersw90,huangl94} the imaginary part was found to appear also for a
non-Abelian like background field. However, it was argued~\cite{kaisersw90}
that the imaginary part originates in the abuse of the formula
$\leff \propto \ln\det G^{-1} $, where $G$ is a propagator,
valid only for positive definite $G$. It was concluded~\cite{kaisersw90}
 that the contribution from the unstable modes and the imaginary part
should not be trusted. In \refc{starinetsvz94} the imaginary part was
shown to vanish for large enough values of a color condensate $A_0$.\bigskip

Since a full treatment of the tachyonic mode is out of the scope of this
monograph, some approximations or assumptions are required.
In for  example \refs{kapusta81,mullerr81} the contribution from the tachyonic
mode
was neglected, as we have done here in Section~\ref{sec-massless}.
However, in \refs{dittrichs81,ninomiyas81}
it was argued that the contribution of the unstable mode should be taken into
account by  analytical continuation, as in \eq{dmassfin}.
Starting from the partition function
or the free energy of the quark gluon plasma, cf. \eq{genfreen}, it seems
very unnatural to include an {\em unphysical} tachyonic mode. Due to the
presence of the thermal distribution function the thermal contribution
of the tachyonic mode will not be purely imaginary, unlike the vacuum
contribution. In \refc{hbnielsenn79} the existence of a zero-energy mode,
after the removal of the tachyonic mode, was pointed out. This suggests that
including  contributions only from $p_z^2+m^2-eB\geq 0$ in the lowest
energy mode is a reasonable approximation.
However, this will exclude the contribution from the plausible
condensate. We shall therefore briefly comment also on the corresponding
results when the real part of the contribution from the
unstable mode is included. These problems  may
be solved by introducing a thermal gluon mass, large enough to remove the
instability.
The effects of such a mass are considered in Section~\ref{sec-gluonmass}.

Usually a temperature dependent (as well as RGE running) coupling is obtained
from the
vertex correction. In the background field formalism here employed the vertex
correction
is related to the vacuum polarization. We shall therefore consider the vacuum
polarization, and review the obtained effective couplings in
Section~\ref{sec-vacpol}.\medskip

The conserved charge in QCD may
correspond to the color charge, the baryon number or the quark flavor.
Each quark is carrying the
color charge ``1'' of some color, and the baryonic number $1/3$ that
relates the baryon chemical potential to the color chemical potential.
Nature only seems to allow for equal amounts of the different colors, that
corresponds
to equal chemical potentials.  However, the gluons carry equal amount of color
and
anti-color (but not the respective), resulting in a vanishing chemical
potential for
them, as well as their linear combination in terms of the $W$ fields.
The $W$ bosons carry  an equal amount of color and anti-color, in the ideal
combination
for interactions with the external field. The problem with Bose--Einstein
condensation in the lowest energy mode  thus never occurs, since $\mu=0$ for
the
gluons and $W$ bosons.
Nevertheless, we shall here mainly focus on the high temperature situation with
vanishing charge density (i.e. $\mu=0$).
 Since a finite chemical potential  only affects  the quarks, we may in this
case immediately use the results obtained for  QED in
Section~\ref{sec-dense}.\\

\Subsection{sec-connect}{Connections with Spinor and Vector QED}
We shall here first  relate  the theory of mass-less
spin $\sigma=1$, gyro-magnetic ratio $\gamma=2$ bosons interacting with
a background magnetic field, to $SU(N_c)$ Yang--Mills theory in a
background chromo-magnetic field (cf. \refs{nielsen81,huang82}). The
Lagrangian of $ SU(N_c)$ Yang--Mills, is
\be{ymlag}
	\cL_{\rm YM}^{N_c} =-\frac14 \sum_{a=1}^{N_c^2-1} G^a_{\mu\nu}
	G^{a\mu\nu}\equiv \sum_{a=1}^{N_c^2-1} \cL^{(a)} ~~~,
\ee
where
\be{gdef}
	 G^a_{\mu\nu}= \del_\mu E^a_\nu -\del_\nu E^a_\mu + g \sum_{b,c}
	f^{abc} E^b_\mu E^c_\nu~~~.
\ee
We shall now consider this theory in the presence of a background
chromo-magnetic  field, that we choose in the $a' \equiv N_c^2-1$ direction
in color space, i.e.
\be{backe}
	E^{N_c^2-1}_\mu \equiv A_\mu + \tilde E^{N_c^2-1}_\mu~~~.
\ee
Here $A_\mu$ is the vector potential corresponding to a static uniform magnetic
field, and $ \tilde E^{N_c^2-1}_\mu$ is the quantum field.
As in the Abelian theories considered before, we are interested in the one-loop
effective Lagrangian for this background magnetic field $A_\mu$.
To the one-loop order we may neglect every occurrence of $g$, when not in the
combination $gA_\mu$. We then find
\be{lyma}
	 \cL^{(a')}= -\frac14 F_{\mu\nu}F^{\mu\nu}-\frac12 g  F^{\mu\nu}
	\sum_{j=1}^{N_c^2-2} \sum_{k=1}^{N_c^2-2} f^{a'jk}
	E^j_\mu E^k_\nu~~~,
\ee
where $ F_{\mu\nu} \equiv \del_\mu A_\nu -\del_\nu A_\mu$ is the (color)
 electro-magnetic field-strength tensor.
As in the Abelian theories considered above we have not explicitly written
out the radiation field $ \tilde E^{N_c^2-1}_\mu$, but we must remember to
include its thermal contribution. In the limit $g=0$ the Lagrangian of
\eq{ymlag} corresponds to $(N_c^2-1)$ ``photons'', that gives the thermal
contribution for $B=0$ to the effective Lagrangian.
 For $j' \neq a'$ we find
\be{lymj}
	\cL^{(j')} = -\frac14 \left[ \del_\mu E^{j'}_\nu -\del_\nu E^{j'}_\mu
	- g \sum_{k=1}^{N_c^2-1} f^{a' j' k} ( A_\mu E^k_\nu -A_\nu E^k_\mu )
	\right]^2~~~,
\ee
where we have used the total anti-symmetry of the structure constants
$f^{j' a' k}= -f^{a' j' k}$. Generally we may choose the generators of
$SU(N_c)$, such that $f^{a' j' k}$ is non-vanishing only for one $k=k'$, for
fixed $a'$ and $j'$. Let us now define
\bea{wdef}
	E^{j'}_\mu &\equiv& \frac1{\sqrt{2}} [ W^{(j',k')}_\mu +
	W^{(j',k')\dagger}_\mu]~~~, \non \\
	E^{k'}_\mu &\equiv& \frac1{i\sqrt{2}} [ W^{(j',k')}_\mu -
	W^{(j',k')\dagger}_\mu]~~~.
\eea
In terms of the charged vector boson field $W$ we have
\be{lymkj}
	\cL^{(j')}+\cL^{(k')} =-\frac12
	 [ D^{*}_\mu   W^{(j',k')\dagger}_\nu -
	 D^{*}_\nu W^{(j',k')\dagger}_\mu ]
	 [ D^\mu  W^{(j',k')\nu} -D^\nu  W^{(j',k')\mu} ]~~~,
\ee
where we have defined the covariant derivative
\be{covder}
	D_\mu W^{(j',k')}_\nu \equiv (\del_\mu -i g f^{a'j'k'} A_\mu)
	 W^{(j',k')}_\nu~~~.
\ee
Adding the relevant term from $\cL^{(a')}$ in \eq{lyma}, we find
\bea{lykjtot}
	\cL^{(j',k')}-\frac14 F^2 &=& -\frac14 F^2
	-\frac12[ D^{*}_\mu   W^{(j',k')\dagger}_\nu -
	 D^{*}_\nu W^{(j',k')\dagger}_\mu ]
	[ D^\mu  W^{(j',k')\nu} - D^\nu  W^{(j',k')\mu} ] \non \\
	&& -ig f^{a'j'k'}F^{\mu\nu} \frac{W_\mu^{(j',k')\dagger}
	W_\nu^{(j',k')}- W_\mu^{(j',k')}  W_\nu^{(j',k')\dagger}}{2}~~~.
\eea
Comparing with \eq{lvec}, we see that this is exactly the Lagrangian for
a spin $\sigma=1$, gyro-magnetic ratio $\gamma=2$ boson of charge
$ -g f^{a'j'k'}$, interacting with a background field, described by $A_\mu$.
The terms in the original Lagrangian $\cL_{\rm YM}^{N_c}$ describing the
self-interaction of the $E^a$ fields, that we here have discarded when
considering the one-loop effective Lagrangian for $A_\nu$, are essential for
the
renormalizability of the theory~\cite{huang82}.
The total Lagrangian $\cL_{\rm YM}^{N_c}$ is then obtained by summing over
all such pairs $(j',k')$
\be{lymtot}
	\cL_{\rm YM}^{N_c}= -\frac14 F^2 + \sum_{(j,k)}\cL^{(j,k)}~~~,
\ee
where $(j,k)$ are the pairs with non-vanishing $f^{a'jk}$, for $a'=N_c^2-1$.
Again using  the total anti-symmetry of the structure constants, the total
charge
squared of these particles is
\be{veccharge}
	e^2_{\rm vec} = g^2 \sum_{(j,k)} (f^{a'jk})^2=
	\frac{g^2 }2  \sum_{j=1}^{N_c^2-1} \sum_{k=1}^{N_c^2-1}
	f^{a'jk} f^{a'jk}~~~.
\ee
Now the $SU(N_c)$ generators in the adjoint representation are
$(T^a_{\rm vec})_{jk}= f^{ajk}$. We thus find a group invariant squared charge,
independent of the direction $a'$ in $SU(N_c)$
\be{vecgentr}
	e^2_{\rm vec}\delta^{ab} =\frac{g^2 }2 \Tr [ T^{a}_{\rm vec}T^{b}_{\rm vec}]=
	\frac{g^2 }2 N_c \delta^{ab}~~~.
\ee
We now wish to couple fermions in the fundamental representation to our
Yang--Mills theory. The corresponding Lagrangian reads
\be{fermym}
	\cL_{\rm YM}^{\rm ferm}=
	 \psibar ( i \del \sslask +g \sum_{a=1}^{N_c^2-1}
	E^a\!\!\slask~\; T_f^a -m ) \psi~~~,
\ee
where the color indices $j=1,2, \ldots , N_c$ have been suppressed.
In the background field $A_\mu$ of \eq{backe}, we again neglect $g$ when not
combined in $gA_\mu$, and find
\be{lagfermym}
	\cL_{\rm YM}^{\rm ferm}= \psibar ( i \del \sslask +g
	A \slask\;  T_{\rm ferm}^{a'} -m ) \psi~~~.
\ee
Generally we may choose $T_{\rm ferm}^{a'}$ diagonal, i.e. $T_{\rm ferm}^{a'}
= \mbox{\rm diag}(
	t_1,t_2, \ldots ,t_{N_c})$. Comparing with \eq{lqed}, we see that
 the Lagrangian in \eq{lagfermym}
corresponds to $N_c$ fermions with charges $e=-g (t_1,t_2, \ldots ,t_{N_c})$,
coupled to the external field described by $A_\mu$. The squared sum of their
charges is then also group invariant
\be{fermcharge}
	e_{\rm ferm}^2\delta^{ab}= \left(g^2 \sum_{j=1}^{N_c} t_j^2\right)\delta^{ab}
	= g^2 \Tr[ T_{\rm ferm}^{a}  T_{\rm ferm}^{b} ]= \frac{g^2}2\delta^{ab}~~~.
\ee
Invariance under background field gauge transformations requires the product
$eB$ to be invariant under renormalization (see e.g. \refc{chaichianh95}).
We may then immediately use the results of the previous Sections.
We shall for simplicity use the relations
between quadratic charges in \eqs{veccharge}{fermcharge}. When terms not
quadratic in the coupling appear in the effective Lagrangian, the correct
approach is to sum over the moduli of the constituent charges to the
power considered, as we assume $eB>0$. However, when not only containing even
powers of
 the coupling $e$, the result is  not group-invariant. This means that
it depends on the specific direction in color space chosen for the magnetic
field.
Any such result is dubious, and probably unphysical.

\Subsection{sec-effqcd}{The Effective Coupling in Thermal QCD}
In the previous Section we related QCD with a background (chromo-)
 magnetic field
to the corresponding cases in spinor and vector QED. However, for the
contributions to the effective Lagrangian independent of the magnetic field,
we must instead identify and sum over all degrees of freedom.
We shall here consider  $N_f$ flavors of quarks with possibly different
masses $m_f$. They all come in $N_c$ different colors. The number of gluons
is $N_c^2-1$. Since we only are summing over the true degrees of freedom,
either in the vacuum energy, or in the free energy, no Faddeev--Popov
 ghosts are needed.
The effective Lagrangian for $SU(N_c)$ Yang-Mills with a background color
magnetic field, and  $N_f$ flavors of quarks in the fundamental representation
may thus be written as
\be{lqcdgen}
	\leffc= \ltre +N_c\sum_f \lmatn^{\rm ferm} + (N_c^2-1)\lmatn^{\rm vec}
	+\sum_f \lmate^{\rm ferm}(e_{\rm ferm}^2= \frac{g^2}2) +
	 \lmate^{\rm vec}(e_{\rm vec}^2=\frac{N_c}2 g^2)~~~,
\ee
with  $m \mapsto m_f$ in the different terms in the sum over flavors.
In vacuum the corresponding renormalization gives  the correct lowest
order QCD $\beta$-function
\be{betaqcd}
	\lam \frac{d g(\lam)}{d \lam}\equiv \beta^{\rm QCD}[g(\lam)]=- \ch^{\rm QCD}
	g^3(\lam)~~~,
\ee
where we obtain from \eqs{cqed}{chivec} and \eqs{vecgentr}{fermcharge}
\be{chiqcd}
	\chc= \frac{N_c}2 \ch^{\rm vec} +\frac{N_f}2 \ch^{\rm ferm} =
	\frac{11 N_c -2 N_f}{48\pi^2}~~~.
\ee
In the high temperature limit $\{ T^2 \gg eB,~m_f^2 \}$ for all flavors of
quarks,  we find when neglecting the contribution from the tachyonic mode
\bea{lqcdtot}
	\leffc&=&\pi^2 T^4 \left\{ \frac{7}{180}N_c N_f + \frac2{45}(N_c^2-1)
	\right\} -\frac12 \frac{(gB)^2}{g^2(\lambda)} -\frac12 \chc (gB)^2 \ln
	\frac{T^2}{\lambda^2} \non \\
	&&  -\frac{1}{2\pi^2} \left( \frac{N_c}2 \right)^{3/4}(gB)^{3/2}T \left[
	\ln\frac{T^2}{4gB} +2 \pi(\sqrt2 -1) |\zeta(-1/2)|
	+2-\pi \right] +\cO[(gB)^2]. \non \\
	&&
\eea
Here   the ordinary renormalization at $T=0, \mu=0$ has been performed at the
renormalization scale $\lam$.
The dependence of $\lam$ in $g(\lam)$ is cancelling the explicit $\lam$
dependence,
so that the effective Lagrangian  is  independent of $\lambda$, as  follows
 from its  definition.
Notice that the $(gB)^2 \ln(m^2)$ terms cancel between \eq{fingenleff} and
\eq{masshight}, so   that there is no
dependence of the different quark masses whatsoever, as long as
$T^2 \gg m_f^2$, and that the same result follows from the mass-less case in
 \eq{lessspin}. Due to the appearance of $(gB)^{3/2}$ it is not quite correct
to use the squared average charge for the gluons. In
 $SU(3)$ with $B$ in the $N_c^2-1=8$ direction in color space,
we have the relevant structure constants $f^{458}=\sqrt3/2=f^{678}$
(see e.g. \refc{huang82}). We should
thus substitute $(N_c/2)^{3/4} \equiv (3/2)^{3/4} \mapsto 2\,(3/4)^{3/4}$
in the term containing
$(gB)^{3/2}T$, but this depends on the direction in color space.
In \refs{kapusta81,mullerr81}, where the tachyonic mode was neglected,
 no high temperature expansion was made, and the
graphs presented do not extend to  temperatures high enough,
 that we could  compare our result with theirs.\bigskip

 Let us now use \eq{aeffdef} to define a field and temperature
dependent effective coupling $\gef(T,eB)$.
  Comparing the
effective charge at different temperature and field strengths, we find
\be{geffla}
	\frac1{\gef^2(T,eB)}\simeq \frac1{\gef^2(T_0,eB_0)}+\chc \ln\frac{T^2}{
	T_0^2}+ F(T/eB)-F(T/eB_0)~~~,
\ee
where we have defined
\be{fdef}
	F(x) \equiv x \left(\frac{N_c}2\right)^{3/4} \frac{3}{4\pi^2} \left[
	\ln\frac{x^2}{4}  -2\pi (\sqrt2 -1)|\zeta(-\frac12)|
	+1-\pi \right]~~~.
\ee
 This effective coupling is thus {\em decreasing}
as a function of the temperature. The leading behavior, with $x\equiv T/
\sqrt{eB}$ is
\be{lead}
	\gef^2(x)\simeq \frac{\gef^2(x_0)}{1+\gef^2(x_0) 3 (N_c/2)^{3/4} x \ln(x)/
	(2\pi^2)}~~~.
\ee
This is a faster decreasing coupling than predicted by the vacuum RGE with
the identification $\lambda \mapsto T$.
The dominant  $T/\sqrt{eB}\, \ln(T^2/eB)$ term is neither present
 when considering the running coupling obtained from the vacuum polarization
 (see Section~\ref{sec-vacpol}), nor when including the real part of the
 contribution from  the tachyonic mode as considered next. In this latter
case we find instead
\be{realvec}
	\lmate^{\rm QCD}+\lvacc = \frac{(gB)^{3/2}T}{2\pi}2\left( \frac34
	\right)^{3/4}[ (1+i) +2(\sqrt2 -1)
	\zeta(-1/2)] -\frac12 \ch^{\rm QCD} (gB)^2
	\ln \frac{T^2}{\lam^2}~~~.
\ee
We have here summed over two vector bosons  with the charge $e=g \sqrt3/2$,
to obtain the  coefficient in front of $(gB)^{3/2}T$ for comparisons with
earlier
results. The factor
$2 (3/4)^{3/4}$  should not be present in $SU(2)$ Yang--Mills theory
(with $f^{123}=1$).

Evaluating  numerically we find $[ 1+2(\sqrt2 -1)\zeta(-1/2)]\simeq  0.827781$,
that exactly equals the corresponding numerical coefficient in
\refc{ninomiyas81} (where $SU(2)$ was considered).
 As a matter of fact, using the same identification with a
generalized Riemann $\zeta$ function as used here, the  coefficient of
\refc{ninomiyas81}
may be written as $i+[2^{3/2}|\zeta(-1/2,3/2)|-1]\simeq i+0.827781$.
However, we do not agree with the high temperature limit in \refc{dittrichs81},
and particularly not with the sign of the imaginary part. That sign of the
imaginary part is obtained if we substitute $m^2-eB \rightarrow i\eps-eB$ (i.e.
opposite to the Feynman prescription in the propagator) of  \eq{dmassfin}.
It was pointed out already in \refc{ninomiyas81} that the different sign on
the imaginary part is unphysical, and corresponds to a blow up instead of a
decay of the corresponding configuration.
In \refc{starinetsvz94} (where $SU(2)$
was considered) a coefficient of $(eB)^{3/2}T/2\pi$ in perfect agreement
with \eq{realvec} is obtained (in the limit $A_0=0$, that is what here is
considered), but the real  contribution from
$\dlmat$ (i.e. ``1'' in $1+i$) is unfortunately lost in the final result.
  Probably a typographic error also
has caused a coefficient in front of the $(gB)^2 \ln(T^2/\lam^2)$ not in
accordance to the RGE in \refc{starinetsvz94}.
Considering the real part, \eq{realvec}
 gives an effective charge that is {\em increasing} with the temperature. The
leading
behavior of $F$ in \eq{geffla} is in this case
\be{leadre}
	F(x)= x\, 2\left( \frac34 \right)^{3/4}\, \frac3{4\pi} [ 1 +2(\sqrt2 -1)
	\zeta(-1/2)]~~~.
\ee
Notice that the effective coupling in \eq{lead} and the one obtained from
\eq{leadre} are not invariant
under transformations in color space. In \eq{leadre} we have explicitly stated
the
result when the magnetic field is chosen in the ``8'' direction in $SU(3)$ in
order
to compare with previous results. We feel dubious to the physical significance
of
such an effective charge.

\Subsection{sec-vacpol}{The Vacuum Polarization}
At finite temperature, the broken Lorentz invariance results in two different
possible tensor structures in the vacuum polarization
\be{vacpol}
	\Pi^{\mu\nu}(k)\equiv P_T^{\mu\nu} \Pi^T(k) + P_L^{\mu\nu} \Pi^L(k) \qquad,
\ee
where $ P_T^{\mu\nu}$ and $P_L^{\mu\nu}$ are the spatially transverse and
longitudinal
polarization operators, respectively.
 Considering the effective coupling obtained from the vacuum polarization for
 a gluon with
momentum scale $|{\bf k}|=\kappa$ in mass-less  thermal QCD, we may write
\be{gluon}
	\frac1{g^2(\kappa,T)}=\frac1{g^2(\kappa_0,T_0)} + \chc \ln \frac{
	\kappa^2}{\kappa_0^2} + \tp(T/\kappa)- \tp(T_0/\kappa_0)~~~.
\ee
We have here already used the well known form of the vacuum polarization in the
absence
of matter, leading to \eq{rensol}.
Chaichian and Hayashi~\cite{chaichianh95} suggest to use either
$g^2\tp=\Pi_{\rm mat}^L/\kappa^2=(k^2/\kappa^2) \Pi_{\rm mat}^{00}/\kappa^2$,
or $g^2\tp=\Pi^T_{\rm mat}/\kappa^2=(\sum_j\Pi^{jj}_{\rm mat}
-(k_0^2/\kappa^2)\Pi^{00}_{\rm mat})/(2\kappa^2)$.
We shall here only consider the static momentum configuration, $k^0=0,~
|{\bf k}|=\kappa$.

With the formalism used in this work we may easily calculate $\Pi_{\rm
mat}^{00}$
in the limit $k_\nu=0$.
Using the similarity between the way the chemical potential $\mu$ and
the vector potential $A_0$  enters the Lagrangian, we find
\be{vacpoltr}
	\pimat(k_\nu=0)=e^2 \frac{\del^2 \lmat}{\del \mu^2}~~~.
\ee
Since we have found the quark masses to be irrelevant in the high temperature
limit, let us for simplicity consider the mass-less case only.
Performing the derivative in \eq{lmat1}, we may then let $\mu=0$. The field
independent part is then obtained as
\be{vacless}
	\Pi^{00}_{\rm mat,0}=\frac{e^2}{2\pi^2} T^2 \sum_s \sum_{l=1}^\infty
	\frac{(-1)^{2\sigma(l-1)}}{l^2} \int_0^\infty \frac{dt}{t^2}\exp[-
	\frac1{2t} ]~~~.
\ee
This gives $\Pi_{\rm mat,0}^{00,{\rm scal}}=
\Pi_{\rm mat,0}^{00,{\rm ferm}}=e^2T^2/3$, and in the vector case
$\Pi_{\rm mat,0}^{00,{\rm vec}}=2e^2T^2/3$.
What amounts to the field dependent part we proceed much in the same way as
in Appendix~\ref{app-less}. Substituting $u=\beta^2l^2eBt/2$, we may
directly perform the Poisson resummation, since the $l=0$ term is vanishing in
this case. Again it is necessary to separate out the $k=0$ term for bosons.
In the scalar case this gives a contribution
\be{knoll}
	\Pi_{\rm mat,1}^{00,{\rm scal}} =
	-\sqrt{eB}T \frac{2-\sqrt2}{4\pi}[-\zeta(1/2)]~~~.
\ee
The other terms are found to be suppressed at high temperatures. Identifying
the momentum scale $\kappa^2=eB$, we thus find in QCD, using for $k_0=0$
  $\Pi_{\rm mat}^L= \Pi^{00}_{\rm mat}$
\be{pimatt}
	\tp(x)=x^2 \left( \frac{N_c}3 +\frac{N_f}6 \right) -x \frac{N_c}2
	\frac{2-\sqrt2}{2\pi}|\zeta(1/2)| +\cO(1/x)~~~.
\ee
The leading $T^2$ behavior agrees with \refc{chaichianh95}, but the
coefficient in front of $x=T/\kappa$, approximately  $
	0.1365 N_c/2$, does not. The absence of $\ln T$ in
\eq{pimatt} indicates that the identification between the magnetic field $eB$
and the momentum scale $\kappa^2$ does not work here.\bigskip

On the other hand, using the gauge invariant
 Vilkovisky--De Witt effective action indicates
that it is the transverse part of the vacuum polarization tensor that
governs the renormalization~\cite{landsman89,elmforsk95}.
In this case a cancelation occurs
between the leading $T^2/\kappa^2$ terms. The coefficient in front of the
next leading $T/\kappa$ term is found to depend on the gauge fixing parameter.
However, the Vilkovisky--De Witt effective action speaks in favor of the
 Landau gauge $\xi=0$. Recently Elmfors and Kobes~\cite{elmforsk95}
 used a Braaten--Pisarski
resummation scheme~\cite{braatenp90} to calculate $\pitr$ self-consistently
in the high-temperature limit. Their result  reads in our notation
\be{perres}
	\tp(x)\simeq - x \frac{N_c}{64}[(3+\xi)^2+14]~~~.
\ee
This will give an effective charge increasing with the temperature. The effects
of the inclusion of a non-perturbative magnetic mass was also considered
 in \refc{elmforsk95}. In the Landau gauge this did not qualitatively
 change the
asymptotic behavior of the effective charge, whereas it could do so in
other gauges with $\xi>1$. Recently, Sasaki~\cite{sasaki95} has used the
pinch-technique~\cite{cornwall82} to calculate a gauge-invariant thermal
$\beta$-function in QCD. It was explicitly shown in various gauges that the
thermal
$\beta$-function is invariant, and the result agrees with the one obtained in
the
background field formalism in the Feynman gauge $\xi=1$. No Braaten--Pisarski
resummation was performed in this case, but we have found no reason why
\eq{perres}
with $\xi=1$ should not be the resummed pinch technique result.
We have no answer why the Vilkovisky--De Witt effective action, on general
grounds
supposed to be gauge-invariant, gives a different result from the pinch
technique,
explicitly shown to give equal results in various gauges.

\Subsection{sec-gluonmass}{Inclusion of the Thermal Gluon  Mass}
It is well-known that the solution of  some infra-red problems in thermal field
theories
requires a resummation of the dominant diagrams to all orders in perturbation
theory~\cite{braatenp90}. The result is concluded in the elegant ``Hard Thermal
Loop
Effective Action''. Here we are primarily interested in regularizing the
infra-red
behavior. We shall therefore  only  take the thermal gluon
mass into account.
The derivation performed here will only be heuristic.
Considering the high temperature limit, we
shall use the well-known result~\cite{braatenp90} for $B=0$
\be{gluonmass}
	m_G^2\equiv \Pi^T(k^2=0)\simeq 
	\left( N_c +\frac{N_f}2 \right)g^2 \frac{T^2}9\qquad,
\ee
that must be the leading  contribution also  for $T^2\gg gB$.
This is the on-shell  self-energy for   transverse gluons only, corresponding to
the polarizations
$s=\pm 1$ for the charged $W$ bosons.
 We thus add and subtract a mass term
$m_G^2/2 \,  \sum_a E_a^\nu E^a_\nu$ to the $SU(N_c)$ Lagrangian in \eq{ymlag}.
The subtracted mass term is  necessary in order not to change the original
Lagrangian,
 that would ruin gauge invariance. It is treated as a counter term, in order
that
the hard thermal loops  are not  counted twice. The resummed transverse gluon
self-energy is for example  $\Pi^T_{\rm res}=\Pi^T-m_G^2$. The contributions
from
this counter term are calculated in Appendix~\ref{app-glumass}.
Rewritten in terms of $W$, the resummed Lagrangian  corresponds to \eq{lvec}
with
$m^2=m_G^2$.
Therefore, we may  immediately  use the results from massive vector QED,
but only with two polarizations $(s=\pm 1)$ similar to the mass-less case,
since a dynamical mass term cannot change
the number of degrees of freedom. Furthermore, we shall assume the thermal
mass to
be so large that the instability is removed, i.e. $m_G^2>gB$.
With these assumptions we may immediately use our previous results in the high
temperature limit  to find
\bea{maglag}
	\lmate^{\rm QCD} + \lvacc& \simeq&  -\frac12 \chc (gB)^2 \ln\frac{T^2}{\lam^2}
	- 2 \frac{T}{m_G} \frac{N_c}2 \frac{(gB)^2}{48\pi^2}\non \\
	&&
	+\sqrt{\frac{N_c}2} \frac{gBT}{2\pi} \left( \sqrt{m_G^2+\sqrt{\frac{N_c}2}gB}-
	\sqrt{m_G^2-\sqrt{\frac{N_c}2}gB}\;\; \right)\quad,
\eea
where $g$ and $B$ denote the charge and field renormalized at the momentum
scale $\lam$,
respectively. Actually we should here also have included the thermal quark
mass
$m_q^2=[(N_c^2-1)/2N_c]g^2 T^2$, but this gives no relevant effects.
Notice that expanding according to $gB<m_G^2$  in the last term of \eq{maglag}
 will only give even  powers of
\be{evenp}
	\left( \frac{\sqrt{\frac{N_c}2}gB}{m_G^2} \right)^{2n} =
	\left( \frac{N_c}2 \frac{(gB)^2}{m_G^4} \right)^n \qquad,
\ee
so that the result is group invariant to any order. The correct approach would
 be to use $e^4_{\rm vec} =g^4 /2 \, \Tr [ T^{a'}_{\rm vec}T^{a'}_{\rm vec}
 T^{a'}_{\rm vec}T^{a'}_{\rm vec} ]$ etc. for higher powers.
Here we shall only consider the leading behavior for $m_G^2\gg gB$. The
contribution from the counter terms is then
\be{contcon}
	\lmate^{\rm QCD,(c)}+ \lvac^{\rm QCD, (c)} \simeq \frac12\chc (gB)^2 \left[
	I_1\left( \frac{m_G}T\right) +1 \right]\qquad.
\ee
 We may now perform the derivative with respect to $B$ to find the
magnetization
\be{magmassmag}
	\frac{M^{\rm QCD}}{g^2B} \simeq -\chc \left[ \ln \frac{T^2}{\lam^2}
	-I_1\left( \frac{m_G}T\right)-1 \right]
	+\frac{N_c}2 \frac{T}{m_G} \frac{12-1}{12\pi}\quad.
\ee
We have here written ``$12-1$'' in the last term
 to indicate that this coefficient enters exactly as
$\ch^{\rm vec} =2\ch^{\rm scal}+\Delta\ch^{\rm vec}$.
In terms of the charge
 renormalized at  momentum scale $\lam_0$, we now find the effective
temperature
dependent charge
\be{leadmagch}
	\frac1{\geff^2(T,gB)} \simeq \frac1{g^2(\lam_0)} +\chc\left[
 \ln \frac{T^2}{\lam_0^2}-I_1\left( \frac{m_G}T\right)-1 \right]
	-\frac{N_c}2 \frac{T}{m_G} \frac{11}{12\pi}\qquad.
\ee
Notice that $m_G \propto T$, so that the only  $T$ dependence is in the
logarithm.
The other constant terms may be renormalized away.
The effective charge obtained with a thermal  mass regularizing the instability
is thus decreasing logarithmically in accordance to the vacuum RGE with
$\lam=T$.
This was the original assumption by Collins and Perry~\cite{collinsp75},
on the formation of a quark--gluon plasma (a notion  later coined) as a result
of
asymptotic freedom. Also notice that the result follows provided that the
thermal mass
is $m_\beta \propto T$, and $m_\beta^2 >gB$, regardless of the
explicit form of $m_\beta$. We could therefore have used the electric mass
$m_E^2=\Pi^L(k_0=0,{\bf k})=3m_G^2$, or the (nonperturbative) magnetic mass
$m_M^2=\Pi^T(k_0=0,{\bf k})=(c g^2 T)^2$. The thermal   mass used here 
follows also in the uniform
limit $ \Pi^T(k_0,{\bf k}={\bf 0})=m_G^2$.
We believe that it is the thermal gluon  mass used here
that is relevant for the physical polarizations of the gluons.

\Section{sec-disc}{Discussion}
In this paper we have calculated the effective charge of different gauge
theories in vacuum and in a  thermal environment. In the vacuum case
the effective charge is related to the magnetic susceptibility.
The general criteria for (anti-) screening, is that the effective charge is
(increasing) decreasing with the distance or the inverse momentum scale.
If the reference charge that we are comparing our effective charge with
is taken as the bare charge, measured at infinite energy, the effective charge
is always smaller than the bare charge in spinor and scalar QED. This
corresponds to screening, and
is related to bare diamagnetism. In QCD, on the other hand, the effective
charge is always greater than the bare charge. This implies anti-screening,
asymptotic freedom and bare paramagnetism. However, in spinor and scalar QED
the reference charge
 is customary taken as the classical charge, measured in the infinite
wave-length limit. In this case the effective charge is always greater than
the reference charge, and the corresponding vacuum magnetization will show a
paramagnetic behavior. These theories do anyhow exhibit screening since
the effective charge is increasing with the energy scale.
In QCD there is no such natural scale at which to define the effective charge,
since the charge is becoming infinitely large in the long wavelength limit.
The magnitude of the effective charge compared to the reference charge
, and thereby also the sign of the magnetic susceptibility, will therefore
depend on the relative magnitude between the scales at which the reference
charge and the effective charge are measured.\bigskip

However, some caution is required in the interpretation of asymptotic freedom
in QCD in terms of anti-screening in a dielectricum. In the presence of
quarks and gluons, {\em external fields are screened } also in QCD, see e.g.
\refc{gyulassy85}. Chromo-electric fields are screened with the electric
(Debye) mass $m_E^2 \simeq g^2[ (N_c+N_f/2)T^2/3+ \sum_f \mu_f^2/(2\pi^2)]$.
To cure some infrared problems also magnetic fields are believed to be screened
with a non-perturbatively generated magnetic mass $m_M^2 = \cO(g^4T^2)$.
We believe that this could be related to the condensate removing the
tachyonic mode, as suggested already by Cornwall~\cite{cornwall82}, but this
needs to
be investigated further. We have here found
that the anti-screening is caused by the large spin magnetic moment of the
gluons that themselves carry the color charge, in terms of the $W$ field in
\eq{wdef}. We may view the anti-screening as an effect of dispersion of the
color charge~\cite{gyulassy85}. For example, a static blue (B) quark may become
red (R) by emitting a B\={R} gluon. This will effectively distribute
the blue charge over a volume $\simeq r^3$. When investigating the charge in a
volume $\lam^{-3} \ll r^3$, only a small fraction of the net blue charge is
found.\bigskip

In the thermal case we mainly focus on the effective charge in QCD. There are
several advantages in this approach of calculating the effective charge
from the effective Lagrangian in a back-ground magnetic field. Background
field gauge-invariance is maintained, even though we only consider one
particular choice of gauge here. This gauge-invariance implies that
the product of the coupling and the background field are invariant under
renormalization. The quark--gluon, three gluon and four gluon couplings are
thus renormalized in the same way, and kept equal. Summing over physical
degrees of freedom only, in the vacuum energy and the free energy, there is
no need to introduce ghosts. The running coupling obtained from the RGE is a
function of scaled momenta $k^\mu \rightarrow e^t k^\mu$, but this means that
we move off-shell. Most often the effective charge is calculated in the static
limit $k^\mu=(0,\k \hat{k})$, relevant for the screening of external fields.
 This means that one is in the deep Euclidean
region $k^2=-\k^2$. The running coupling so obtained cannot be ascribed to
real particles, only to virtual particles in internal processes. The effective
coupling obtained from the effective Lagrangian, on the other hand, is
directly related to the interactions of the real particles in the heat and
charge-bath. Moreover, the thermal contribution to the effective Lagrangian
is related to the free energy. Since physical, measurable quantities are
obtained by performing derivatives on the free energy,
 it must be gauge invariant and cannot depend on the gauge fixing,
 up to an irrelevant constant.\bigskip

 At fixed magnetic field,
 we find from \eq{geffla} the leading behavior of the
effective coupling in QCD as a function of the temperature
\be{gefft}
	\gef^2(T,gB)=\frac{\gef^2(T_0,gB)}{1+\gef^2(T_0,gB) \frac{3}{2\pi^2}
	\left(\frac{N_c}2\right)^{3/4} \left[
	\frac{T}{\sqrt{gB}} \ln\frac{T}{\sqrt{gB}} -\frac{T_0}{\sqrt{gB}}
	\ln\frac{T_0}{\sqrt{gB}} \right]}~~~.
\ee
As found when explicitly calculating parts of the vacuum polarization tensor,
we cannot transform the dependence on the magnetic
field, to dependence of momentum through $\k^2 \approx eB$. However, we may
consider the limit of vanishing magnetic field.
In the limit $gB \rightarrow 0$ in \eq{gefft} we find
the simple behavior
\be{tlin}
\frac{\gef^2(T)}{\gef^2(T_0)}=\frac{T_0}T~~~,
\ee
i.e. a coupling {\em linearly decreasing} with the temperature. The effective
charge
obtained in the limit of vanishing magnetic field has the advantage of being
independent
of the direction in color space for the magnetic field, unlike the effective
charge in \eq{gefft} as indicated by the non-analytical behavior in terms of
$N_c$.
Notice, however,
 that the appearance of the $(gB)^{3/2}T\ln(T/\lambda)$ term  in the
effective Lagrangian, that enforces asymptotic freedom at high temperatures,
is solely due to the negligence of the tachyonic mode.
 If this mode is
taken into account, by analytically continue $m^2 \rightarrow -i \eps$
in \eq{dmassfin}, this term disappears.  But we are in that case left with an
imaginary part in the free energy. Considering the real part only we find
in the high temperature limit an {\em increasing} effective charge
\be{htincleff}
	\gef^2(T,gB)=\frac{\gef^2(T_0,gB)}{1-\gef^2(T_0,gB) \frac{3}{4\pi}
	\left(\frac{N_c}2\right)^{3/4} \left[
	\frac{T}{\sqrt{gB}}  -\frac{T_0}{\sqrt{gB}} \right]}~~~.
\ee
Also this effective charge suffers from non-invariance under transformations in
 the
color  group $SU(N_c)$, and thus cannot be trusted.
In this case we cannot take the limit $gB \rightarrow 0$, since this would
take us across the Landau pole, at which the coupling is becoming infinitely
strong.
 We obtain qualitatively different results
whether the tachyonic mode is
neglected, or the real part of its contribution is taken into account.
However,
 particularly when starting from the free energy,  we consider it
physically more reasonable to neglect all contributions from the imaginary
energy mode, that of course cannot be present in a real situation.
The group invariant charge obtained in the field free limit when neglecting the
contribution from the tachyonic mode could thus be taken as an indication of
the true situation.\medskip

A thorough investigation of this tachyonic
mode has so far required numerical treatment, and the (small) change
in the magnetic field due to the condensate in this mode would also alter
the other modes present, and thereby also the effective Lagrangian.
This seems to destroy the advantage of simplicity in obtaining
the effective charge, using the effective Lagrangian in a magnetic field.
However, including a thermal gluon  mass may  take us out of this dilemma.
We find in the high temperature limit when the thermal  mass is large enough
to completely remove the instability, an effective charge {\em decreasing
according to
the vacuum RGE}~(\ref{rensol}) with $\lam \mapsto T$
\be{rgeffch}
	\frac1{\geff^2(T)}=\frac1{\geff^2(T_0)}  +\chc \ln \left(\frac{T}{T_0}
	\right)^2 \qquad.
\ee
This is what naively has been expected for long~\cite{collinsp75}, but hitherto
not
obtainable. Obviously, this effective charge  is also valid in the absence of a
(chromo-) magnetic field.
However, the derivation outlined here is merely heuristic, and more
rigorous considerations are to be  be performed.\bigskip

If we instead consider the case of large flavor chemical potentials, we may
immediately use the expansions in \eq{highmu} and \eq{moscmu}. It is the
latter, oscillating contribution that is most interesting. It is
oscillating with an amplitude $|M_{\rm osc}|/e^2B \propto \mu/\sqrt{eB}$,
but it may be positive as well as negative depending on the ratio
$(\mu^2-m^2)/(2eB)$. The effective coupling thus may increase as well as
decrease. However, due to these rapid oscillations we are doubtful to this
definition of the effective charge in the limit of large chemical potentials
and low temperatures.\vspace{13mm}

\begin{center}
{\bf ACKNOWLEDGMENTS}\\ \end{center}
This paper is an extension of previous work that I have done in cooperation
with  Per Elmfors, Bo-Sture Skagerstam, and Per
Liljenberg. Special thanks to P.~E. for supplying me with
his imaginary time calculations,
and to B.-S.~S. for reading the manuscript. I am also grateful to Joakim
Hallin and B.-S.~S. for enlightening discussions about the
 group invariance; and to Maurice Jacob for his remark about thermal
masses. Some useful hints for the finalization of this study were obtained
during
the network meeting supported by NorFA grant no. 96.15.053-O. \bigskip

\bc
	{\bf \Large  APPENDIX}
\ec

\appendix
\Section{app}{High Temperature Expansions}

\Subsection{app-deltmass}{Massive Vector QED}
We shall here find the high temperature limit of
\be{deltmass2}
	\dlmat=\frac{eB}{\pi^2} \int_0^\infty dp_z p_z^2 \left[
	 \frac1{E_{0,-1}} \frac1{e^{\beta E_{0,-1}}-1} - \frac1{E_{0,0}}
	\frac1{e^{\beta E_{0,0}}-1} \right]~~~.
\ee
Using the expansion of the distribution function
\be{distexp}
	\frac1{e^{\beta E}- (-1)^{2\sigma}} = \sum_{l=1}^\infty
	(-1)^{2\sigma(l-1)} e^{-\beta E l}~~~,
\ee
and the identity
\be{smartid}
	\frac{ e^{-\beta E l}}{ \beta E l} \equiv
	\int_0^\infty \frac{dt}{\sqrt{2\pi t}} \exp\left[ -\frac12 \left(
	\beta^2 E^2 l^2  t +\frac1t \right) \right]~~~,
\ee
we may perform the Gaussian $p_z$-integral in \eq{deltmass}. We can also
identify~\cite{gradshteyn}
\be{besselk}
	\int_0^\infty \frac{dt}{t^{\nu+1}} \exp\left[-bt -\frac{c}t \right]=
	2 \left(\frac{b}c \right)^{\nu/2} K_\nu(2\sqrt{bc})~~~,
\ee
where $ K_\nu=K_{-\nu}$ is a modified Bessel function. We may now write
\be{delmass}
	\dlmat= \frac{eBT^2}{\pi^2}\left[ I\left(\frac{\sqrt{m^2-eB}}T \right)
	- I\left(\frac{\sqrt{m^2+eB}}T \right) \right]~~~,
\ee
where
\be{idef}
	I(x) \equiv \sum_{l=1}^\infty \frac1{l^2} (xl) K_1(xl)~~~.
\ee
Using the derivative property of Bessel functions~\cite{hmf}, we
find~\cite{gradshteyn}
\be{ider}
	\left( \frac1x \frac{d}{dx}\right) I(x)=-\sum_{l=1}^\infty K_0(xl)=
	-\left\{ \frac12 \left( \gamma_E + \ln \frac{x}{4\pi} \right)
	-\frac\pi{2x} +\cO(x)\right\}~~~.
\ee
This finally gives
\bea{dmassfin2}
	\dlmat& =& -\frac{(eB)^2}{4\pi^2} \ln \left[ \frac{T^2(4\pi)^2}{\sqrt{
	m^4-(eB)^2}}\right] +\frac{eBT}{2\pi} (\sqrt{m^2+eB}- \sqrt{m^2-eB} )
	 \non \\
	&& + \frac{eB m^2}{8\pi^2} \ln \left( \frac{m^2+eB}{m^2-eB} \right)
	-\frac{(eB)^2}{2\pi^2} \left( \frac12 -\gamma_E \right)
	+eBT^2 \cO\left(\frac{m^2+eB}{T^2} \right)^{3/2}~~~.
\eea
In order for this to be valid also for $eB>m^2$, we must consider the
analytical
continuation defined by $m^2 \rightarrow m^2-i\eps$. The same result is
obtained
also if we treat the contribution for $p_z^2<eB$ separately in analogy to the 
vacuum case.
The result is presented in \eq{dmassfin}.

\Subsection{app-less}{Mass-less Spinor and Scalar QED}
In the mass-less case, we have found it necessary to add the vacuum
contribution
in order to perform a Poisson resummation and find the high temperature
behavior. We have not managed to perform this analysis using the cut-off
regularization previously utilized. Therefore,
 we shall in this section instead
use a dimensional regularization in $4-2\delta$ dimensions. In order to keep
the coupling dimension-less, we substitute $e \mapsto \lambda^\delta e$, where
$\lambda$ is an energy scale. For $E_{n,s}>\mu$ we may use the expansion in
\eq{distexp}, and the identity in \eq{smartid}. Performing the summation
over $n$, we find
\bea{lmat1}
	\lmat&=& \frac{eB\lambda^\delta}{2\pi} \sum_s \int_0^\infty \frac{dt}{
	\sqrt{2\pi t}} \sum_{l=1}^\infty (-1)^{2\sigma(l-1)} \beta l
	 \cosh(\beta l \mu) \int\left( \frac{dp_z}{2\pi} \right)^{1-2\delta}
	 p_z^2 \exp\left(
	-\frac1{2t}\right) \non \\
	&& \times  \frac{ \exp[- \beta^2 l^2 t ( p_z^2 - eB\lambda^\delta
	\gamma s)/2]}{\sinh(\beta^2 l^2 e B \lambda^\delta t/2)}~~~.
\eea
Integrating  over $p_z$, subtracting the $B=0$ part $\lmatn$ and substituting
$u=\beta^2 l^2 eB \lambda^\delta t/2$, we find
\bea{lmatett}
	\lmate&=& \frac{(eB)^2}{8\pi^2} \left( \frac{4\pi \lambda^2}{eB}
	\right)^\delta
	 \sum_{l=1}^\infty (-1)^{2\sigma(l-1)} \cosh(\beta l \mu)
	\int_0^\infty \frac{du}{u^{3-\delta}}\exp\left(- \frac{ \beta^2 l^2 eB
	\lambda^\delta}{4u} \right) \non \\
	&& \times \sum_s  \left\{ \frac{u \exp(\gamma s u)}{ \sinh(u)}
	-1 \right\}~~~.
\eea
We now wish to perform a Poisson resummation, and must therefore add the $l=0$
term to extend  the summation  to $\sum_{l=-\infty}^\infty$. This term
with $l=0$ is the bare vacuum Lagrangian, regularized in $4-2\delta$
dimensions,
cf. \eq{lvacgenfin}.
Let us now for simplicity consider the case of a neutral plasma, $\mu=0$.
Again the vector case needs special care, so here we first consider only
spinor and scalar QED with $\sigma=1/2$ and $\sigma=0$, respectively.
Performing the Poisson resummation we find
\bea{lmatvac}
	\lvac + \lmate& =& (eB)^2 \frac{2T\sqrt{\pi}}{\sqrt{eB\lambda^\delta}}
	\left(\frac{4\pi \lambda^2}{eB} \right)^\delta \sum_{k=-\infty}^\infty
	\int_0^\infty \frac{du}{u^{5/2-\delta}} \exp\left[ -\frac{4\pi^2 T^2}{
	eB\lambda^\delta} (k+\sigma)^2u \right] \non \\
	&& \times \frac{(-1)^{2\sigma}}{16\pi^2} \sum_s \left\{
	\frac{u\exp(\gamma s u)}{\sinh[u]} -1 \right\}~~~.
\eea
The term with $k=0$ for bosons ($\sigma=0$) gives a contribution to $\lmats$,
finite for $\delta=0$
\be{matbos}
	 \frac{(eB)^{3/2}T}{8\pi^{3/2}}\int_0^\infty
	\frac{du}{u^{5/2}}  \left\{
	\frac{u}{\sinh[u]} -1 \right\} =  \frac{(eB)^{3/2}T}{8\pi^{3/2}}
	(2^{5/2}-4)\sqrt{\pi}
	\zeta(-\frac12)~~~,
\ee
where $\zeta(-1/2)\simeq -0.207886 $ ,
 and we have used $\Gamma(-1/2)=-2\sqrt\pi$.
 We have here identified Riemann's zeta function
$\zeta(z)$ from an integral
representation in \refc{gradshteyn} that is valid for \mbox{$\Re( z) >1$}.
 However the
subtracted ``1'' on the left hand side of \eq{matbos}
is regularizing this expression (the integral is manifestly
convergent), and we have numerically checked the equality.
For the other terms in the sum, we sum over $s$ and expand~\cite{gradshteyn}
\be{expansion}
	 \frac{(-1)^{2\sigma}}{8\pi^2}  \left\{
	\frac{u\cosh(\gamma \sigma u)}{\sinh[u]} -1 \right\}= \frac12 \ch u^2 -
	\frac1{8\pi^2}\sum_{n=2}^\infty \frac{ 2^{2n}-2(1-\gamma\sigma)}{(2n)!}
	B_{2n}u^{2n}~~~,
\ee
for $\gamma\sigma=0,~1$.
We may then integrate over u.
Due to the alternating sign of the Bernoulli numbers $B_{2n}$,
 the modulus of the sum is smaller
than the modulus of the first neglected term. For $\nu \geq 2$ we may let
$\delta=0$, and find the contribution
\be{highernu}
	-\frac1{8\pi^2}\frac{ 2^{2n}-2(1-\gamma\sigma)}{(2n)!} B_{2n}
	 (eB)^2 \left(\frac{eB}{4\pi^2 T^2} \right)^{2(\nu-1)}
	\frac{\Gamma(2\nu-3/2)}{\sqrt{\pi}} \zeta(4\nu-3, 1-\sigma)~~~,
\ee
that is suppressed for large $T$.
{}From
the first term in the expansion of \eq{expansion} we obtain a contribution
\be{lettsum}
	\frac12 \ch (eB)^2 \left( \frac{\lambda^2}{\pi T^2} \right)^\delta
	\frac{\Gamma(\frac12+\delta)}{\sqrt{\pi}} {\sum_{k}}' \frac1{|k+\sigma|^{1+
	2\delta}}~~~,
\ee
where $\sum'_k $ means that the term with $k=0$ should be excluded for $\sigma
=0$. We can now identify a generalized Riemann's
$\zeta$-function~\cite{gradshteyn}
\be{zetafun}
	 {\sum_k}' \frac1{|k+\sigma|^{1+2\delta}}\equiv 2\sum_{k=0}^\infty
	\frac1{[k+(1-\sigma)]^{1+2\delta}}\equiv 2\zeta(1+2\delta,1-\sigma)~~~.
\ee
As $\delta \rightarrow 0$ we have the expansion~\cite{gradshteyn}
\be{zetaexp}
	 \zeta(1+2\delta,1-\sigma)= \frac1{2\delta} +\gamma +2\sigma \ln 2 +
	\cO(\delta)~~~,
\ee
for $\sigma=0,1/2$. We  now expand $\Gamma(\frac12+\delta)=\sqrt{\pi} +
\cO(\delta)$, $(\lambda^2/\pi T^2)^\delta=1-\delta \ln(\pi T^2/\lambda^2) +
	\cO(\delta)$,  perform a renormalization to absorb the
divergent term $1/\delta$ and all terms $\propto (eB)^2$ in the bare coupling,
 and then let $\delta$ vanish.

\Subsection{app-deltless}{Mass-less Vector  QED}
In this case we cannot employ a
technique similar to \eq{matbos}, since the integral would become divergent
due to the $\exp(\gamma s u)$. Here we shall
 neglect the tachyonic mode and start integrating at $p_z^2=eB-m^2$ in the
contribution from the lowest Landau level.
Since the lowest energy then
is vanishing, only $\mu=0$ is possible in this case.
If we have a finite charge density it must therefore reside in a
Bose-Einstein condensate in the zero energy mode.
When the tachyonic mode is not included, integrations by part will cause
non-vanishing surface terms. We shall here start from the
 free energy in  \eq{genfreen}, that we believe is most physical. We then have
\be{deltless}
	\dlmat=-\frac{eBT}{\pi^2}\left\{ \int_{\sqrt{eB}}^\infty dp_z
	\ln[ 1- e^{-\beta E_{0,1}}] -\int_0^\infty dp_z
	\ln[ 1- e^{-\beta E_{0,0}}] \right\}~~~.
\ee
Substitute $x=\beta E_{0,s}$, and split the integral at 
$1 \gg x_0 \gg \sqrt{eB}/T$. Expanding for $x>x_0$
\be{sqrtexp}
	\frac1{\sqrt{x^2+eB/T^2}}-\frac1{\sqrt{x^2-eB/T^2}} =
	\frac{eB}{T^2} \frac1{x^3} + \cO\left( \frac{eB}{T^2} \right)^3~~~.
\ee
The remaining integral is convergent, so this will only produce an
irrelevant $\cO (eB)^2$
term, and terms suppressed at large $T$. For $x<x_0$ we instead expand
\be{logexp}
	\ln[1-e^{-x}]=\ln x - \frac{x}2 +\cO(x^2)~~~.
\ee
We may now perform the integrals over $x$ to find \eq{deltlessfin}.

\Section{app-glumass}{Thermal Gluon Mass}
We shall here briefly outline the results from the inclusion of a thermal mass
$m_\beta$.
 Let us start without the matter contribution.
Considering one-loop effects only, we may integrate
 out the particles in the generating functional of Green's functions.
Adding and subtracting the term corresponding to the thermal mass, we
 find in the simplest case of a mass-less (i.e. without the thermal mass)
scalar field
\be{logtr}
	\int d^4 x \leff= \int d^4 x \left( -\frac14 F^2 \right) + i\, \Tr\ln[ i
	(-D^2-m^2_\beta+m^2_\beta)]\quad.
\ee
We now wish to define our perturbative expansion in terms of the massive field.
Let us therefore split the logarithm and the effective Lagrangian according to
\be{logsplit}
	\int d^4 x \left( \lvac + \lvac^{(c)} \right) =i\,\Tr\ln[i(-D^2-m^2_\beta)] +
	i\, \Tr\ln\left[ 1+\frac{m^2_\beta}{-D^2-m^2_\beta} \right]\quad.
\ee
Performing the $m^2_\beta$ derivative on the first term, we recognize the
propagator
$(-D^2-m^2_\beta)^{-1}$. This may then be generalized to arbitrary spin,
resulting in \eq{genvac} for the case of a magnetic field considered here.
Expanding the second logarithm in \eq{logsplit} to leading order, we again find
the
trace of the propagator. Similar generalizations to arbitrary spin give
\be{vaccon}
	\lvac^{(c)} \simeq m^2_\beta \, i(-1)^{2\sigma}
	\frac{eB}{(2\pi)^3}
	 \sum_{s}  \sum_{n=0}^\infty  \int
	d \om  d  p_z \frac{1}{w^2-E_{n,s}^2+ i\eps}~~~.
\ee
Integrating over $\om$, we may then use \eq{sqrtid}. In this counter term we
shall
encounter no divergences, so we may immediately let $1/\Lambda=0$. We may then
perform the Gaussian integral in $p_z$, and sum the infinite geometrical series
in $n$.
Substituting $t=x^2$, and subtracting the contribution for $B=0$, we find
\be{convacgen}
	\lvac^{(c)}\simeq m^2_\beta \frac{(-1)^{2\sigma}}{16\pi^2} \sum_s
	\int_{0}^\infty
	\frac{dt}{t^2} \exp(-m^2_\beta t) \left\{ \frac{eBt}{\sinh(eBt)}
	\exp(eBt \gamma s )-1 \right\}~~~.
\ee
Expanding for $eBt \simleq eB/m^2_\beta \ll 1$, we find
\be{canvacfin}
	\lvac^{(c)}\simeq \frac12 \ch (eB)^2 \left[ 1+ \cO \left(\frac{eB}{m^2_\beta}
	\right)^2 \right]\quad.
\ee

Let us now consider the thermal contribution. Similar generalizations as in the
vacuum
case leads for  $\mu=0$  to
\be{conterm}
	\lmat^{(c)} \simeq m^2_\beta \frac{eB}{(2\pi)^2} \sum_s \sum_{n=0}^\infty \int
dp_z
	\frac{1}{E_{n,s}} \frac1{e^{\beta E_{n,s}} -
	(-1)^{2\sigma}}~~~.
\ee
Expanding the distribution function according to \eq{distexp}, and using
\eq{smartid},
we may again perform the Gaussian integral in $p_z$, and sum the infinite
geometrical
series in $n$. Subtracting the contribution for $B=0$, we find
\bea{matconfin}
	\lmate^{(c)}& = &\frac{(-1)^{2\sigma}}{8\pi^2} \sum_s \sum_{l=1}^\infty
	(-1)^{2\sigma l} \frac1{\frac12 \beta^2 m^2_\beta l^2} \int_0^\infty
	\frac{dt}{t^2}
	\exp\left[ -\frac12 \beta^2 m^2_\beta l^2 t -\frac1{2t} \right] \non \\
	&& \times\left\{ \frac{\frac12 \beta^2 l^2 eB t}{\sinh(\frac12
	\beta^2 l^2 eB t)}
	 \exp(\frac12 \beta^2 l^2 eB\gamma s  t) -1 \right\}\quad.
\eea
We may again expand according to $eB\beta^2 l^2 t/2 \simleq eB/m^2_\beta \ll
1$. Using
\eq{besselk} the result reads
\be{contermres}
	\lmate^{(c)}\simeq \frac12 \ch (eB)^2 \left[ I_\sigma(m_\beta/T) +
	\cO(eB/m^2_\beta) \right]\quad,
\ee
where we have defined
\be{isigdef}
	 I_\sigma(x)\equiv 2 \sum_{l=1}^\infty (-1)^{2\sigma l} xl K_1(xl)\quad.
\ee
For small $z$, $zK_1(z) \approx 1$, and for large $z$, $z K_1(z) \approx
\sqrt{\pi z/2}
	\exp(-z)$~\cite{hmf}, so that the sum is convergent. Now, to leading order,
$m_\beta=m_G \propto T$, so $I_\sigma(m_\beta/T)$ is $T$-independent.
For completeness we may also write down the field independent contribution
\be{conin}
	\lmatn{(c)} =m^4 \frac{(-1)^{2\sigma}}{2\pi^2}\sum_s \sum_{l=1}^\infty
	\frac{(-1)^{2\sigma l}}{\beta m  l} K_1(\beta m l)\quad.
\ee
We have found no analytical expression for this series, but its leading
behavior is
$\cO(m^2_\beta T^2)$, i.e. $ \cO(g^2 T^4)$, sub-leading for small couplings.
Of course, this field independent part is irrelevant for the effective
coupling.
%
\jump

\end{document}